\documentclass[12pt]{article}
\usepackage{amsfonts}
\usepackage[dvips]{graphicx}
\usepackage{latexsym,amsmath,amssymb,graphics,stmaryrd}
\usepackage{cite}
\usepackage{array}
\usepackage{subfigure}
\usepackage{multirow}
\usepackage{epsfig}
\usepackage[dvips]{graphicx}
\usepackage[dvips]{color}
\usepackage{pstricks}
\usepackage{pst-node}
\usepackage{pst-plot}
\usepackage{dsfont}
\usepackage{amsthm}
\usepackage{graphics}
\usepackage{subfigure}
\usepackage{fancyhdr}
\usepackage[dvips]{color}

\usepackage{feynmp}

\fancypagestyle{plain}{
\fancyhf{}
\newcommand{\fpage}{\iffloatpage{}{\thepage}}
\fancyfoot[C]{\fpage}

}

\newcommand{\col}{~,}
\newcommand{\pnt}{~.}
\newcommand{\AdS}{\text{AdS}}

\newcommand{\la}{\lambda}
\newcommand{\hla}{\hat{\lambda}}
\newcommand{\bl}{\bar\lambda}
\newcommand{\NN}{\mathcal{N}}
\newcommand{\s}{\sigma}

\newcommand{\unitmatrix}{\mathds{1}}
\newcommand{\comm}[2]{\left[#1\smash[b]{\mathbin{,}}#2\right]}
\newcommand{\acomm}[2]{\left\{#1\smash[b]{\mathbin{,}}#2\right\}}

\newcommand{\de}{\operatorname{d}\!}

\newcommand{\e}{\operatorname{e}}

\newcommand{\pfour}[4]{{}\{#1,#2,#3,#4\}{}}
\newcommand{\pthree}[3]{{}\{#1,#2,#3\}{}}
\newcommand{\ptwo}[2]{{}\{#1,#2\}{}}
\newcommand{\pone}[1]{{}\{#1\}{}}

\newlength{\neglength}
\newlength{\diameter}
    
\newcommand{\dslash}[1]{
  \ensuremath{
  \text{$\settowidth{\diameter}{$#1$}%
  #1%
  \hspace{-0.5\diameter}%
  \makebox[0pt][c]{\slash}%
  \hspace{0.5\diameter}$}}}

\newcommand{\svertex}[2]{%
\fmfiequ{#1}{point length(#2)/2 of #2}
}
\newcommand{\dvertex}[3]{%
\fmfiequ{#1}{point length(#3)/3 of #3}
\fmfiequ{#2}{point 2length(#3)/3 of #3}
}
\newcommand{\vvertex}[3]{%
\fmfipath{px}
\fmfiequ{px}{(0,ypart(#2))..(100,ypart(#2))}
\fmfiequ{#1}{point xpart(#3 intersectiontimes px) of #3}
}

\newcommand{\vsix}[6]{%
\fmf{plain,tension=1,left=0.25}{#1,vc6}
\fmf{plain,tension=1}{vc6,#2}
\fmf{plain,tension=1,right=0.25}{#3,vc6}
\fmf{plain,tension=1,right=0.25}{vc6,#4}
\fmf{plain,tension=1}{#5,vc6}
\fmf{plain,tension=1,left=0.25}{vc6,#6}
\fmfposition
\fmfipath{p[]}
\fmfipair{vm[],vo[],vi[]}
\fmfiset{p16}{vpath(__#1,__vc6)}
\fmfiset{p26}{vpath(__vc6,__#2)}
\fmfiset{p36}{vpath(__#3,__vc6)}
\fmfiset{p46}{vpath(__vc6,__#4)}
\fmfiset{p56}{vpath(__#5,__vc6)}
\fmfiset{p66}{vpath(__vc6,__#6)}
\svertex{vm1}{p16}
\dvertex{vo1}{vi1}{p16}
\svertex{vm2}{p26}
\dvertex{vi2}{vo2}{p26}
\svertex{vm3}{p36}
\dvertex{vo3}{vi3}{p36}
\svertex{vm4}{p46}
\dvertex{vi4}{vo4}{p46}
\svertex{vm5}{p56}
\dvertex{vo5}{vi5}{p56}
\svertex{vm6}{p66}
\dvertex{vi6}{vo6}{p66}
}

\newcommand{\vacpol}[2]{%
\fmfcmd{
begingroup;
save t,v,tv,do,di,ppol,pstr,dia;
path ppol,pstr;
pair v[],tv[],do[],di[];
ppol=vpath(__#1,__#2);
t1=arctime (1/3*arclength ppol) of ppol;
t2=arctime (2/3*arclength ppol) of ppol;
v1=point t1 of ppol;
v2=point t2 of ppol;
pstr=v1--v2;
t3=arctime (0.5*arclength pstr) of pstr;
v3=point t3 of pstr;
dia=arclength pstr; 
fill(fullcircle scaled dia shifted v3) withcolor 0.2black;
endgroup;
}
}

\newcommand{\nnint}[2]{%
\fmfcmd{
begingroup;
save t,v,tv,do,di,pstr,dia;
path pstr;
pair v[],tv[],do[],di[];
pstr=#1--#2;
t3=arctime (0.5*arclength pstr) of pstr;
v3=point t3 of pstr;
tv3=unitvector direction t3 of pstr;
dia=arclength pstr; 
fill(fullcircle scaled dia yscaled 0.5 rotated angle(tv3) shifted v3) withcolor 0.2black;
endgroup;
}
}

\newcommand{\plainwrap}[4]{%
\fmfipath{pi[]}
\fmfiset{pi1}{vloc(__#1) ..controls (-0.175w,ypart(vloc(__#1))) and (-0.175w,-0.15w) .. (xpart(vloc(__#2)),-0.15w)}
\fmfiset{pi2}{(xpart(vloc(__#2)),-0.15w) ..(xpart(vloc(__#3)),-0.15w)}

\fmfiset{pi3}{(xpart(vloc(__#3)),-0.15w) ..controls (1.175w,-0.15w) and (1.175w,ypart(vloc(__#4))) .. vloc(__#4)}
\fmfi{plain}{pi1 ..pi2 ..pi3}
}
\newcommand{\wigglywrap}[4]{%
\fmfipath{pi[]}
\fmfiset{pi1}{#1 ..controls (-0.175w,ypart(#1)) and (-0.175w,-0.15w) .. (xpart(vloc(__#2)),-0.15w)}
\fmfiset{pi2}{(xpart(vloc(__#2)),-0.15w) ..(xpart(vloc(__#3)),-0.15w)}

\fmfiset{pi3}{(xpart(vloc(__#3)),-0.15w) ..controls (1.175w,-0.15w) and (1.175w,ypart(#4)) .. #4}
\fmfi{photon}{pi1}
\fmfi{photon}{pi2}
\fmfi{photon}{pi3}
}
\newcommand{\dasheswrap}[4]{%
\fmfipath{pi[]}
\fmfiset{pi1}{vloc(__#1) ..controls (-0.175w,ypart(vloc(__#1))) and (-0.175w,-0.15w) .. (xpart(vloc(__#2)),-0.15w)}
\fmfiset{pi2}{(xpart(vloc(__#2)),-0.15w) ..(xpart(vloc(__#3)),-0.15w)}
\fmfiset{pi3}{(xpart(vloc(__#3)),-0.15w) ..controls (1.175w,-0.15w) and (1.175w,ypart(vloc(__#4))) .. vloc(__#4)}
\fmfiset{pi4}{pi1 ..pi2 ..pi3}
\fmfi{dashes}{pi1}
\fmfi{dashes}{pi2}
\fmfi{dashes}{pi3}
}

\newcommand{\vsixrangefourl}{%
\fmftop{v3}
\fmfbottom{v4}
\fmfforce{(0.125w,h)}{v3}
\fmfforce{(0.125w,0)}{v4}
\fmffixed{(0.25w,0)}{v1,veu1}
\fmffixed{(0.25w,0)}{v2,v1}
\fmffixed{(0.25w,0)}{v3,v2}
\fmffixed{(0.25w,0)}{v4,v5}
\fmffixed{(0.25w,0)}{v5,v6}
\fmffixed{(0.25w,0)}{v6,ved1}
\fmf{plain}{ved1,veu1}
\vsix{v1}{v2}{v3}{v4}{v5}{v6}
\fmfipair{ve,veo[],vem[],vei[]}
\fmfipath{pe}
\fmfiset{pe}{vpath(__ved1,__veu1)}
\vvertex{veo1}{vo1}{pe}
\vvertex{vem1}{vm1}{pe}
\vvertex{vei1}{vi1}{pe}
\vvertex{veo6}{vo6}{pe}
\vvertex{vem6}{vm6}{pe}
\vvertex{vei6}{vi6}{pe}
\svertex{ve}{pe}
}

\newcommand{\ftrianglerangethree}{%
\fmftop{v3}
\fmfbottom{v4}
\fmfforce{(0.125w,h)}{v3}
\fmfforce{(0.125w,0)}{v4}
\fmffixed{(0.25w,0)}{v2,v1}
\fmffixed{(0.25w,0)}{v3,v2}
\fmffixed{(0.25w,0)}{v4,v5}
\fmffixed{(0.25w,0)}{v5,v6}
\fmffixed{(0,0.433h)}{vt3,vt1}
\fmffixed{(0.125w,whatever)}{vt1,v2}
\fmfpoly{phantom}{vt1,vt3,vt2}
\fmf{dashes}{vt1,vi1}
\fmf{dashes}{vi1,vc1}
\fmf{dashes}{vc1,vo1}
\fmf{dashes}{vo1,vt2}
\fmf{dashes}{vt2,vi2}
\fmf{dashes}{vi2,vc2}
\fmf{dashes}{vc2,vo2}
\fmf{dashes}{vo2,vt3}
\fmf{dashes}{vt3,vi3}
\fmf{dashes}{vi3,vc3}
\fmf{dashes}{vc3,vo3}
\fmf{dashes}{vo3,vt1}
\fmf{plain,right=0.125}{vt2,v1}
\fmf{plain,left=0.25}{v2,vt1}
\fmf{plain,left=0.25}{vt1,v3}
\fmf{plain,left=0.25}{v4,vt3}
\fmf{plain,left=0.25}{vt3,v5}
\fmf{plain,right=0.125}{v6,vt2}
\fmffreeze
\fmfposition
\fmfposition
\fmfipath{pt[],p[]}
\fmfipair{vt[],vm[],vi[],vo[]}
\fmfiset{p4}{vpath(__v4,__vt3)}
\fmfiset{p5}{vpath(__vt3,__v5)}
\fmfiset{p6}{vpath(__v6,__vt2)}
\svertex{vm4}{p4}
\svertex{vm5}{p5}
\svertex{vm6}{p6}
}

\DeclareMathOperator{\tr}{tr}

\DeclareMathOperator{\diag}{diag}

\DeclareMathOperator{\perm}{P}

\numberwithin{equation}{section}
\newlength{\eqoff}

\newlength{\unit}
\psset{xunit=\unit,yunit=\unit,runit=\unit}
\newlength{\linew}
\psset{linewidth=\linew}
\begin{document}
\begin{fmffile}{CSlettergraphs}

\fmfcmd{%
def getmid(suffix p) =
  pair p.mid[], p.off[], p.dir[];
  for i=0 upto 36:
    p.dir[i] = dir(5*i);
    p.mid[i]+p.off[i] = directionpoint p.dir[i] of p;
    p.mid[i]-p.off[i] = directionpoint -p.dir[i] of p;
  endfor
enddef;
}

\fmfcmd{%
marksize=2mm;
def draw_mark(expr p,a) =
  begingroup
    save t,tip,dma,dmb; pair tip,dma,dmb;
    t=arctime a of p;
    tip =marksize*unitvector direction t of p;
    dma =marksize*unitvector direction t of p rotated -45;
    dmb =marksize*unitvector direction t of p rotated 45;
    linejoin:=beveled;
    draw (-.5dma.. .5tip-- -.5dmb) shifted point t of p;
  endgroup
enddef;
style_def derplain expr p =
    save amid;
    amid=.5*arclength p;
    draw_mark(p, amid);
    draw p;
enddef;
def draw_marks(expr p,a) =
  begingroup
    save t,tip,dma,dmb,dmo; pair tip,dma,dmb,dmo;
    t=arctime a of p;
    tip =marksize*unitvector direction t of p;
    dma =marksize*unitvector direction t of p rotated -45;
    dmb =marksize*unitvector direction t of p rotated 45;
    dmo =marksize*unitvector direction t of p rotated 90;
    linejoin:=beveled;
    draw (-.5dma.. .5tip-- -.5dmb) shifted point t of p withcolor 0white;
    draw (-.5dmo.. .5dmo) shifted point t of p;
  endgroup
enddef;
style_def derplains expr p =
    save amid;
    amid=.5*arclength p;
    draw_marks(p, amid);
    draw p;
enddef;
def draw_markss(expr p,a) =
  begingroup
    save t,tip,dma,dmb,dmo; pair tip,dma,dmb,dmo;
    t=arctime a of p;
    tip =marksize*unitvector direction t of p;
    dma =marksize*unitvector direction t of p rotated -45;
    dmb =marksize*unitvector direction t of p rotated 45;
    dmo =marksize*unitvector direction t of p rotated 90;
    linejoin:=beveled;
    draw (-.5dma.. .5tip-- -.5dmb) shifted point t of p withcolor 0white;
    draw (-.5dmo.. .5dmo) shifted point arctime a+0.25 mm of p of p;
    draw (-.5dmo.. .5dmo) shifted point arctime a-0.25 mm of p of p;
  endgroup
enddef;
style_def derplainss expr p =
    save amid;
    amid=.5*arclength p;
    draw_markss(p, amid);
    draw p;
enddef;
style_def dblderplains expr p =
    save amidm;
    save amidp;
    amidm=.5*arclength p-0.75mm;
    amidp=.5*arclength p+0.75mm;
    draw_mark(p, amidm);
    draw_marks(p, amidp);
    draw p;
enddef;
style_def dblderplainss expr p =
    save amidm;
    save amidp;
    amidm=.5*arclength p-0.75mm;
    amidp=.5*arclength p+0.75mm;
    draw_mark(p, amidm);
    draw_markss(p, amidp);
    draw p;
enddef;
style_def dblderplainsss expr p =
    save amidm;
    save amidp;
    amidm=.5*arclength p-0.75mm;
    amidp=.5*arclength p+0.75mm;
    draw_marks(p, amidm);
    draw_markss(p, amidp);
    draw p;
enddef;
}

%
%

\fmfcmd{%
thin := 1pt; 
thick := 2thin;
arrow_len := 4mm;
arrow_ang := 15;
curly_len := 3mm;
dash_len := 1.75mm; 
dot_len := 1mm; 
wiggly_len := 2mm; 
wiggly_slope := 60;
zigzag_len := 2mm;
zigzag_width := 2thick;
decor_size := 5mm;
dot_size := 2thick;
}

\newcommand{\threelthreeg}{
\fmfipair{v[],ve[]}
\fmfipath{ls[]}
\fmfipair{a[]}
\fmftop{vt}
\fmfbottom{vb}
\fmffixed{(0,0.1h)}{vo,vt1}
\fmf{phantom}{vt1,vt}
\fmf{phantom}{vb,vb1}
\fmffixed{(0,0.1h)}{vb1,vi}
\fmffixed{(0,0.75h)}{vi,vo}
\fmf{phantom,right=0.5}{vi,vo}
\fmf{phantom,right=0.5}{vo,vi}
\fmf{phantom}{vi,v0}
\fmf{phantom}{v0,vo}
\fmffreeze
\fmfposition
\fmfiset{ls1}{vpath(__vo,__vi)}
\fmfiset{ls2}{vpath(__vi,__vo)}
\fmfiequ{v3}{point length(ls2)/2 of ls2}
\fmfiset{ls3}{v3--vloc(__v0)}
\fmfiset{ls4}{vloc(__v0)--(vloc(__v0) shifted (100 unitvector direction 1 of ls3 rotated -60))}
\fmfiset{ls5}{vloc(__v0)--(vloc(__v0) shifted (100 unitvector direction 1 of ls3 rotated 60))}
\fmfiequ{a1}{ls1 intersectiontimes ls4}
\fmfiequ{a2}{ls1 intersectiontimes ls5}
\fmfiequ{v1}{point xpart(a1) of ls1}
\fmfiequ{v2}{point xpart(a2) of ls1}
\fmfiequ{v3}{point length(ls2)/2 of ls2}
\fmfiequ{ve1}{v1+(0,0.1h)}
\fmfiequ{ve2}{v2-(0,0.1h)}
\fmfiequ{ve3}{v3+(0.1h,0)}
\fmffreeze
}


\thispagestyle{empty}
\begin{flushright}\footnotesize
\texttt{UUITP-20/09} 
\vspace{0.8cm}\end{flushright}

\renewcommand{\thefootnote}{\fnsymbol{footnote}}
\setcounter{footnote}{0}

\begin{center}
{\Large\textbf{\mathversion{bold} Magnon dispersion to four loops   \\
in the ABJM and ABJ models}}

\vspace{1.5cm}

\textrm{J.~A.~Minahan$^{1}$, O.~Ohlsson Sax$^{1}$ and
C.~Sieg$^{2}$}
\vspace{8mm}

\textit{$^{1}$ Department of Physics and Astronomy, Uppsala University\\
SE-751 08 Uppsala, Sweden}\\
\texttt{joseph.minahan@fysast.uu.se, olof.ohlsson-sax@physics.uu.se} \vspace{3mm}

\textit{$^{2}$The Niels Bohr International Academy\\ The Niels Bohr Institute\\
Blegdamsvej 17,
 DK-2100, Copenhagen \O, Denmark }\\
\texttt{csieg@nbi.dk}
\vspace{3mm}


\par\vspace{1cm}

\textbf{Abstract} \vspace{5mm}

\begin{minipage}{14cm}

The ABJM model is a superconformal Chern-Simons theory with
$\mathcal{N}=6$ supersymmetry which  is believed to be integrable in
the planar limit.  However, there is a coupling dependent function
that appears in the magnon dispersion relation and the asymptotic
Bethe ansatz that is only known to leading order at strong and weak
coupling. We compute this 
function to four loops in perturbation
theory  by an explicit Feynman diagram calculation
for both the  ABJM model and the ABJ extension. 
We find that all coefficients have maximal transcendentality.
We then
compute the four-loop wrapping correction for a scalar operator in the
{\bf 20} of $SU(4)$ and find that it agrees with a recent prediction 
from the ABJM 
$Y$-system of Gromov, Kazakov and Vieira. We
also propose a limit of the ABJ model that might be perturbatively
integrable at all loop orders but has a short range Hamiltonian.  

\end{minipage}

\end{center}

\vspace{0.5cm}


\newpage
\setcounter{page}{1}
\renewcommand{\thefootnote}{\arabic{footnote}}
\setcounter{footnote}{0}

\section{Introduction and summary of results}

There are two seemingly different superconformal field theories which
have an integrable structure in their planar limit.   The first of
these is $\mathcal{N}=4$ super Yang-Mills which has been the subject of
intense study for more than a decade.  The latter is the ABJM model
for a superconformal Chern-Simons theory with gauge group $U(N)\times
U(N)$, where the first gauge group has a Chern-Simons action at level
$k$ and the second has level $-k$ \cite{Aharony:2008ug}.  This theory
was only proposed last year as the conjectured holographic dual to
$M$-theory on $\text{AdS}_4\times\text{S}^7/\text{Z}_k$. 

The integrability of these theories appears in the two point functions
of gauge invariant operators.  In the $\mathcal{N}=4$ case a class of gauge
invariant operators are single traces of $SU(N)$ adjoint fields, whose
anomalous dimensions can be found by mapping the problem to a spin
chain \cite{Minahan:2002ve,Beisert:2003tq,Beisert:2003yb}.  The spins
are in  the singleton representation of $PSU(2,2|4)$, the full
superconformal group of $\mathcal{N}=4$ SYM.  In the ABJM case, the
fundamental fields are in bifundamental representations of $U(N)\times
U(N)$ and the corresponding spin chain is of alternating type
\cite{Minahan:2008hf,Gaiotto:2008cg,Bak:2008cp}. The spins on the odd
sites are in one of the two singleton representations of $OSp(6|4)$,
the full superconformal group of the ABJM model, and the spins on the
even sites are in the other singleton representation
\cite{Zwiebel:2009vb,Minahan:2009te}. 

The spin chains in both theories have ground states that correspond to
the chiral primary operators in their respective gauge theories.
These are operators whose dimensions are protected by the
supersymmetry and thus have zero anomalous dimension.  A convenient
choice for a spin chain ground state in $\mathcal{N}=4$ is  $\tr(Z^L)$, while
in the ABJM model a convenient choice is $\tr((Y^1Y^\dag_4)^L)$.  The
other single trace operators are then constructed by introducing
magnons that change the fields in the chain.  The magnons themselves
transform  in a short representation of a subgroup of the
superconformal group, which is $SU(2|2)\ltimes SU(2|2)$ for $\mathcal{N}=4$
and $SU(2|2)$ for ABJM. 

The presence of the $SU(2|2)$ structures imposes severe constraints on
the magnon dispersion relations. As was shown in
\cite{Beisert:2005tm}, this dispersion relation must have the form 
\begin{equation}
\begin{aligned}\label{E}
E(p)=\sqrt{Q^2+4h^2(\lambda)\sin^2\tfrac{p}{2}}-Q\,,
\end{aligned}
\end{equation}
where $p$ is the momentum of the magnon on the spin chain and $Q$ is
the $R$-charge of the magnon.  For $\mathcal{N}=4$ the charge for a
fundamental magnon is $Q=1$, while in ABJM the charge is $Q=\tfrac{1}{2}$.  
$h^2(\lambda)$ is a function of the 't Hooft coupling $\lambda$, which
in $\mathcal{N}=4$ is $\lambda=g_\text{YM}^2N$, while in ABJM it is
$\lambda=N/k$.  Moreover, the 't Hooft coupling only enters the
asymptotic Bethe equations through this same function $h^2(\lambda)$
\cite{Gromov:2008qe}. 
 
To the best of our knowledge, integrability makes no prediction for
$h^2(\lambda)$, but there exist some alternative analyses not
based on integrability \cite{Berenstein:2008dc,Berenstein:2009qd}. 
Thus, it is important to understand its origin and how
it plays a role in the $\AdS/\text{CFT}$  correspondence.   In the
$\mathcal{N}=4$ case, all  known results in perturbation theory and
from the study of the string theory dual are consistent with
$h^2(\lambda)=\lambda/(4\pi^2)$.  
However, in the ABJM model it is
known at the level of two-loop perturbation theory that
$h^2(\lambda)=\lambda^2+{\rm O}(\lambda^4)$
\cite{Minahan:2008hf,Gaiotto:2008cg,Grignani:2008is}, while at large
coupling we know by taking the BMN limit of type IIA string theory on
$\text{AdS}_4\times\text{CP}^3$ \cite{Nishioka:2008gz,Gaiotto:2008cg} and from one-loop string corrections\footnote{We thank Arkady Tseytlin for comments on this.} \cite{McLoughlin:2008he} that $h^2(\lambda)=\tfrac{1}{2}\lambda -\frac{\ln2}{\sqrt{2}\pi} \sqrt{\la}+ {\rm O}(1)$. In fact, since the
perturbative expansion is even in $\lambda$, the function $h^2(\lambda)$
should have a square root branch cut along the negative real axis.

 
The calculation we present here also considers the ABJ 
modification of the ABJM model
\cite{Aharony:2008gk}, where the gauge group is generalized to
$U(M)\times U(N)$, but with the levels of the gauge group kept at $k$
and $-k$.  In this case there are now two 't Hooft couplings 
defined as
\begin{equation}
\lambda=\frac{M}{k}\col\qquad\hat\lambda=\frac{N}{k}\col
\end{equation}
but the superconformal group is still
$OSp(6|4)$.  It was shown at the two-loop level in the scalar
sector \cite{Bak:2008vd} and in the full $OSp(6|4)$ sector
\cite{Minahan:2009te} that  $\lambda^2$ is replaced by 
$\lambda\hat\lambda$, but
otherwise the dilatation operator is the same, and is therefore still
integrable.\footnote{In the ABJM case, hints for a breakdown of
  integrability beyond the planar limit have explicitly been found in 
\cite{Kristjansen:2008ib}.}
If in the planar limit the integrability is to persist to higher loop
orders, then the only modification that can occur is in the function
$h^2$.
In general it will be a two parameter function,
$h^2(\bar\lambda,\sigma)$, 
where we define $\bar\la$ and $\s$ as
\begin{equation}\label{barlambdasigmadef}
\bar\lambda=\sqrt{\lambda\hat\lambda}\col\qquad
\sigma=\frac{\lambda-\hat\lambda}{\bar\lambda}\pnt
\end{equation}
Since the number of colors is different, parity is
explicitly broken in the ABJ model.  Therefore,  one might expect, for
example, the odd site  magnons would have different dispersion
relations from the even site magnons, or that 
$h^2(\bar\lambda,\sigma)\ne h^2(\bar\lambda,-\sigma)$.  
This breaking has indeed been observed beyond the planar 
limit \cite{Caputa:2009ug}.
However, in the planar limit this does not seem to be the
case. Instead,
there is still a parity symmetry at the level of the planar diagrams 
that prevents either of these from occurring.   Nonetheless, we still find
some interesting behaviour. 

Up to four loops, the expansion 
of $h^2(\bar\lambda,\sigma)$ assumes the following form
\begin{equation}\label{h4ansatz}
\begin{aligned}
h^2(\bar\lambda,\sigma)=\bar\lambda^2+\bar\lambda^4h_4(\sigma)\col\qquad
h_4(\sigma)=h_4+\sigma^2h_{4,\sigma}
\col
\end{aligned}
\end{equation}
where $h_4$ is the same constant as in the ABJM case, in which $\sigma=0$.
In this paper we determine $h_4(\sigma)$ by computing the four-loop 
contribution to the dilatation operator.
We find that the coefficients in 
\eqref{h4ansatz} are explicitly\footnote{ A previous version of this
  paper had a different value of $h_{4,\sigma}$ because of an
  incorrect coupling dependence of $S_5$ in equation (3.6). This
  change only concerns the result in the more general ABJ case.}
\begin{equation}
\begin{aligned}\label{h4result}
h_4=-4\,\zeta(2)\approx -6.58\col\qquad 
h_{4,\sigma}=-\zeta(2)\pnt
\end{aligned}
\end{equation}
The negative sign for $h_4$ is sensible, as this will dampen
the quadratic behaviour found at small $\bar\la$ to the linear behaviour
at large $\bar\lambda$.
Interestingly, $h_4$ has no rational contributions, leaving a maximally 
transcendental result. It is also striking that the coefficients in 
\eqref{h4result} are integers. 

In order to find $h_4(\sigma)$, it was necessary to compute the contributions
of several dozen Feynman diagrams.  The diagrams were evaluated using
dimensional reduction \cite{Chen:1992ee}.  All diagrams started in three
dimensions, where there appear three dimensional
$\epsilon$-tensors from the gauge propagators in the Chern-Simons
Lagrangian and from parity violating fermion terms. 
However, all terms carried an even number of
$\epsilon$-tensors which could be contracted with each other, leaving
metric tensors only.  Once the integrands were in this form, the
integrals were dimensionally regularized, reducing the dimension to
$D=3-2\varepsilon$.

The underlying loop integrals were computed using 
scalar $G$-functions, integration by parts relations 
\cite{Kazakov:1983ns,Broadhurst:1985vq} and generalizations thereof.
Special care had to be taken due to the presence of 
infrared divergences, which in three dimensions arise whenever there are
 three scalar propagators attached to a vertex with  no external
momentum entering the vertex. As checks for the results for the
integrals, we verified relations between integrals and recalculated
certain 
integrals using  GPXT \cite{Chetyrkin:1980pr,Kotikov:1995cw} and also 
certain transformations between two-dimensional master integrals
\cite{Vasiliev:1981dg}.

The result for $h_4$ in (\ref{h4result}) is strikingly similar to the
leading wrapping correction for length four operators, which 
has been obtained by Gromov, Kazakov and Vieira \cite{Gromov:2009tv} 
by means of the proposed $Y$-system 
\cite{Gromov:2009bc,Bombardelli:2009ns,Arutyunov:2009ur} for the 
ABJM model.\footnote{This result is not altered by a correction 
\cite{Bombardelli:2009xz,Gromov:2009at} of the originally proposed $Y$-system.}
In this paper we explicitly compute the
wrapping correction with our methods 
and find agreement with their result. 
 
In section 2 we give  further details on the relation of
$h^2(\bl,\s)$ to the dilatation operator.   
In section 3 we discuss the necessary Feynman diagrams and list their
values.  A more explicit description of our methods will be given in a
later publication \cite{Minahan:2009wg}.
In section 4 we collect  our results from the previous section and
compute $h_4(\s)$.  We also discuss an interesting limit of the ABJ
model where one finds an integrable Hamiltonian with next-to-nearest
neighbour interactions. In section 5 we
compute the wrapping corrections for the length four operator in the {\bf 20}
representation of $SU(4)$.  In section 6 we present our conclusions
and suggest further lines of inquiry.  
There are also three appendices that contain further technical details.

\section{Extraction of $h^2(\bar\lambda,\s)$}

To compute the coefficients $h_4$ and $h_{4,\sigma}$ in a field theory 
calculation, it is
only necessary to consider a single scalar magnon.  Thus, we may
restrict our study to the $SU(2)$ sector of the full $OSp(6|4)$ group.
Actually, we will enlarge this to the $SU(2)\times SU(2)$ sector.
Inside the $SU(2)\times SU(2)$ sector, the magnons  in the first
$SU(2)$ live on the odd
sites, while those in the other $SU(2)$ live
on the even sites.  
The two different types of magnons do not interact
with each other until the six-loop level \cite{Gromov:2008qe} so for
our purposes we may treat them as noninteracting.  

The dispersion relation \eqref{E} yields the following expansion 
in a power series in $\bar\lambda$
\begin{equation}\label{Eexpansion}
\begin{aligned}
E(p)
&=2\bar\lambda^2(1-\cos p)
+2\bar\lambda^4(h_4(\sigma)-3+(4-h_4(\sigma))\cos p-\cos2p)
+\mathcal{O}(\bar\lambda^6)
\col
\end{aligned}
\end{equation}
where we have set $Q=1/2$ and we have reexpressed powers of 
$\sin\frac{p}{2}$ in terms of cosine.
The exponentials which are 
contained in the cosine functions are powers of $e^{ip}$, which 
is the eigenvalue of the shift operator that moves the magnon over by
two sites on the spin chain.  Hence, at the four-loop level there is a
maximal shifting of 4 sites, which does not depend on $h_4$.  However,
for shifts of 2 sites, there is an $h_4$ dependence.  

An ansatz for the dilatation operator which eigenvalues are $E(p)$
can be constructed by considering 
the (open) single magnon momentum / shift operator eigenstate
\begin{equation}\label{onemagnonstate}
\psi_p=\sum_{k=0}^L
e^{ipk}
(Y^1Y^\dagger_4)^kY^2Y^\dagger_4(Y^1Y^\dagger_4)^{L-k-1}
\pnt
\end{equation}
The momentum dependence in the expansion \eqref{Eexpansion}
is a result of acting with certain products of permutations on 
the above state, where a permutation exchanges the fields between
two nearest neighbour either odd or even sites.
Similar to the $\mathcal{N}=4$ SYM case, 
we introduce the permutation structures
\begin{equation}\label{permstruc}
\begin{aligned}
\pfour{a_1}{a_2}{\dots}{a_m}=\sum_{i=1}^L\perm_{2i+a_1\,2i+a_1+2}\perm_{2i+a_2\,2i+a_2+2}\dots\perm_{2i+a_m\,2i+a_m+2}\col
\end{aligned}
\end{equation}
where we identify $L+i\simeq i$ when we act on a cyclic state of length
$L$. Some details about these structures and the permutation basis
at four loops can be found in appendix \ref{app:permstruc}.
Up to four loops, the dilatation operator expands as
\begin{equation}
D=L+\bar\lambda^2D_2+\bar\lambda^4D_4(\sigma)+\mathcal{O}(\bar\lambda^6)
\pnt
\end{equation}
At these orders, $D_k$, $k=2,4$ decomposes into a direct sum 
$D_k=D_{k,\text{odd}}+D_{k,\text{even}}$ of two 
operators, that act non-trivially on either odd or even sites only.
Since up to this order the magnon dispersion relations for 
odd and even sites are identical, the operators for even and odd sites
have identical coefficients and read
\begin{equation}\label{D4}
\begin{aligned}
D_{2,\text{even}}&=\pone{}-\pone{1}\col\\
D_{2,\text{odd}}&=\pone{}-\pone{2}\col\\
D_{4,\text{odd}}(\sigma)
&=(h_4(\sigma)-4)\pone{}+(6-h_4(\sigma))\pone{1}-\ptwo{1}{3}-\ptwo{3}{1}
\col\\
D_{4,\text{even}}(\sigma)
&=(h_4(\sigma)-4)\pone{}+(6-h_4(\sigma))\pone{2}-\ptwo{2}{4}-\ptwo{4}{2}
\pnt
\end{aligned}
\end{equation}
The above expressions are easily obtained by considering 
the action of the permutation structures for odd sites on the state
\eqref{onemagnonstate}. Neglecting boundary effect, we find
\begin{equation}
\begin{aligned}
\pone{}\to 1\col\quad
\pone{1}\to\e^{ip}+\e^{-ip}\col\quad
\ptwo{1}{3}\to\e^{2ip}+2\e^{-ip}\col\quad
\ptwo{3}{1}\to\e^{-2ip}+2\e^{ip}
\pnt
\end{aligned}
\end{equation}
Comparing the resulting expression with \eqref{Eexpansion} then
fixes the coefficients of the non-trivial permutations in
$D_{4,\text{odd}}$ in terms of $h_4(\sigma)$. The coefficient of the
identity $\pone{}$ is then found by demanding that the cyclic ground
state has eigenvalue zero. Since the
dispersion relations for magnons on odd and even sides are identical, 
the same coefficients are found for $D_{4,\text{even}}$.

Notice that in \eqref{D4} there is no term that mixes odd and even
sites, i.e. which 
contains the permutation structure $\ptwo{1}{2}$.
It could in principle appear in a four site interaction, but
explicitly cancels \cite{Bak:2009mq}.

\section{Four-loop calculation}
\label{sec:fourloopcalc}

In the following we list in brief the sets of logarithmically divergent 
diagrams which have to be
considered to compute the asymptotic four-loop dilatation operator $D_4$ 
in the $SU(2)\times SU(2)$ subsector. 
The composite operators in 
this sector do not contain subtraces. We therefore can neglect all
diagrams which only contribute to the coefficient of the trace operator.
Furthermore the contribution of the identity can be inferred from 
supersymmetry. We can hence reconstruct $D_4$ in the $SU(2)\times
SU(2)$ by only considering the diagrams which contribute to flavour 
permutations, and we drop all contributions to the trace and identity 
operator in the result. This simplification is indicated by an arrow
pointing from the diagram
to the respective result.

In all four-loop diagrams, the first leg is attached to an odd site,
i.e.\ to a field $Y^i$ of
the composite operator (denoted by a bold line). The color loop in the 
lower left corner of each diagram will hence
contribute a factor $M$ to the diagram. 
The elementary scalar, fermion gauge and ghost fields
are denoted by respectively plain, dashed, wiggly and dotted lines.
The action from which the Feynman rules are extracted is given in appendix
\ref{app:conventions}.

\subsection{Substructures}

Before turning to the four-loop diagrams, we introduce fixed combinations of 
substructures which appear in the four-loop diagrams.
These substructures themselves do not lead to non-trivial flavour exchanges.
Besides these structures, a four-loop diagram which contributes to the 
coefficients of non-trivial permutations in the 
dilatation operator \eqref{D4} then has to contain a scalar
six vertex. Power counting then restricts the substructures to the
ones shown below.

Flavour-neutral next-to-nearest neighbour interactions 
can only be realized with gauge boson exchange and at least two cubic gauge 
scalar vertices at the first and last leg. The relevant diagrams 
involve the structure
\begin{equation}\label{nnnint}
\begin{aligned}
\settoheight{\eqoff}{$\times$}%
\setlength{\eqoff}{0.5\eqoff}%
\addtolength{\eqoff}{-3.75\unitlength}%
\raisebox{\eqoff}{%
\fmfframe(0,0)(0,0){%
\begin{fmfchar*}(10,7.5)
\fmftop{v1}
\fmfbottom{v4}
\fmfforce{(0.0625w,h)}{v1}
\fmfforce{(0.0625w,0)}{v4}
\fmffixed{(0.4375w,0)}{v1,v2}
\fmffixed{(0.4375w,0)}{v2,v3}
\fmffixed{(0.4375w,0)}{v4,v5}
\fmffixed{(0.4375w,0)}{v5,v6}
\fmf{plain}{v1,vc1}
\fmf{plain}{vc1,v4}
\fmf{plain}{v2,v5}
\fmf{plain}{v3,vc2}
\fmf{plain}{vc2,v6}
\fmffreeze
\fmfposition
\nnint{vloc(__vc1)}{vloc(__vc2)}
\end{fmfchar*}}}
&{}={}
\settoheight{\eqoff}{$\times$}%
\setlength{\eqoff}{0.5\eqoff}%
\addtolength{\eqoff}{-3.75\unitlength}%
\raisebox{\eqoff}{%
\fmfframe(0,0)(0,0){%
\begin{fmfchar*}(10,7.5)
\fmftop{v1}
\fmfbottom{v4}
\fmfforce{(0.0625w,h)}{v1}
\fmfforce{(0.0625w,0)}{v4}
\fmffixed{(0.4375w,0)}{v1,v2}
\fmffixed{(0.4375w,0)}{v2,v3}
\fmffixed{(0.4375w,0)}{v4,v5}
\fmffixed{(0.4375w,0)}{v5,v6}
\fmf{plain}{v1,vc1}
\fmf{plain}{vc1,v4}
\fmf{plain}{v2,vc2}
\fmf{plain}{vc2,v5}
\fmf{plain}{v3,vc3}
\fmf{plain}{vc3,v6}
\fmffreeze
\fmfposition
\fmf{photon}{vc1,vc2}
\fmf{photon}{vc2,vc3}
\end{fmfchar*}}}
{}+{}
\settoheight{\eqoff}{$\times$}%
\setlength{\eqoff}{0.5\eqoff}%
\addtolength{\eqoff}{-3.75\unitlength}%
\raisebox{\eqoff}{%
\fmfframe(0,0)(0,0){%
\begin{fmfchar*}(10,7.5)
\fmftop{v1}
\fmfbottom{v4}
\fmfforce{(0.0625w,h)}{v1}
\fmfforce{(0.0625w,0)}{v4}
\fmffixed{(0.4375w,0)}{v1,v2}
\fmffixed{(0.4375w,0)}{v2,v3}
\fmffixed{(0.4375w,0)}{v4,v5}
\fmffixed{(0.4375w,0)}{v5,v6}
\fmf{plain}{v1,vu1}
\fmf{plain}{vu1,vd1}
\fmf{plain}{vd1,v4}
\fmf{plain}{v2,vu2}
\fmf{plain}{vu2,vd2}
\fmf{plain}{vd2,v5}
\fmf{plain}{v3,vu3}
\fmf{plain}{vu3,vd3}
\fmf{plain}{vd3,v6}
\fmffreeze
\fmfposition
\fmf{photon}{vu1,vu2}
\fmf{photon}{vd2,vd3}
\end{fmfchar*}}}
{}+{}
\settoheight{\eqoff}{$\times$}%
\setlength{\eqoff}{0.5\eqoff}%
\addtolength{\eqoff}{-3.75\unitlength}%
\raisebox{\eqoff}{%
\fmfframe(0,0)(0,0){%
\begin{fmfchar*}(10,7.5)
\fmftop{v1}
\fmfbottom{v4}
\fmfforce{(0.0625w,h)}{v1}
\fmfforce{(0.0625w,0)}{v4}
\fmffixed{(0.4375w,0)}{v1,v2}
\fmffixed{(0.4375w,0)}{v2,v3}
\fmffixed{(0.4375w,0)}{v4,v5}
\fmffixed{(0.4375w,0)}{v5,v6}
\fmf{plain}{v1,vu1}
\fmf{plain}{vu1,vd1}
\fmf{plain}{vd1,v4}
\fmf{plain}{v2,vu2}
\fmf{plain}{vu2,vd2}
\fmf{plain}{vd2,v5}
\fmf{plain}{v3,vu3}
\fmf{plain}{vu3,vd3}
\fmf{plain}{vd3,v6}
\fmffreeze
\fmfposition
\fmf{photon}{vd1,vd2}
\fmf{photon}{vu2,vu3}
\end{fmfchar*}}}
\pnt
\end{aligned}
\end{equation}

Flavour-neutral nearest-neighbour interactions 
involve the following diagrams
\begin{equation}\label{nnint}
\begin{aligned}
\settoheight{\eqoff}{$\times$}%
\setlength{\eqoff}{0.5\eqoff}%
\addtolength{\eqoff}{-3.75\unitlength}%
\raisebox{\eqoff}{%
\fmfframe(0,0)(0,0){%
\begin{fmfchar*}(10,7.5)
\fmftop{v1}
\fmfbottom{v3}
\fmfforce{(0.0625w,h)}{v1}
\fmfforce{(0.0625w,0)}{v3}
\fmffixed{(0.875w,0)}{v1,v2}
\fmffixed{(0.875w,0)}{v3,v4}
\fmf{plain}{v1,vc1}
\fmf{plain}{vc1,v3}
\fmf{plain}{v2,vc2}
\fmf{plain}{vc2,v4}
\fmffreeze
\fmfposition
\nnint{vloc(__vc1)}{vloc(__vc2)}
\end{fmfchar*}}}
&{}={}
\settoheight{\eqoff}{$\times$}%
\setlength{\eqoff}{0.5\eqoff}%
\addtolength{\eqoff}{-3.75\unitlength}%
\raisebox{\eqoff}{%
\fmfframe(0,0)(0,0){%
\begin{fmfchar*}(10,7.5)
\fmftop{v1}
\fmfbottom{v3}
\fmfforce{(0.0625w,h)}{v1}
\fmfforce{(0.0625w,0)}{v3}
\fmffixed{(0.875w,0)}{v1,v2}
\fmffixed{(0.875w,0)}{v3,v4}
\fmf{plain,right=0.125}{v1,vc1}
\fmf{plain,right=0.125}{vc1,v2}
\fmf{plain,left=0.125}{v3,vc2}
\fmf{plain,left=0.125}{vc2,v4}
\fmf{phantom}{vc1,vc2}
\fmffreeze
\fmf{dashes,left=0.5}{vc1,vc2}
\fmf{dashes,left=0.5}{vc2,vc1}
\end{fmfchar*}}}
{}+{}
\settoheight{\eqoff}{$\times$}%
\setlength{\eqoff}{0.5\eqoff}%
\addtolength{\eqoff}{-3.75\unitlength}%
\raisebox{\eqoff}{%
\fmfframe(0,0)(0,0){%
\begin{fmfchar*}(10,7.5)
\fmftop{v1}
\fmfbottom{v3}
\fmfforce{(0.0625w,h)}{v1}
\fmfforce{(0.0625w,0)}{v3}
\fmffixed{(0.875w,0)}{v1,v2}
\fmffixed{(0.875w,0)}{v3,v4}
\fmf{plain}{v1,vc1}
\fmf{plain}{vc1,v3}
\fmf{plain}{v2,vc2}
\fmf{plain}{vc2,v4}
\fmffreeze
\fmf{photon,left=0.25}{vc1,vc2}
\fmf{photon,left=0.25}{vc2,vc1}
\end{fmfchar*}}}
{}+{}
\settoheight{\eqoff}{$\times$}%
\setlength{\eqoff}{0.5\eqoff}%
\addtolength{\eqoff}{-3.75\unitlength}%
\raisebox{\eqoff}{%
\fmfframe(0,0)(0,0){%
\begin{fmfchar*}(10,7.5)
\fmftop{v1}
\fmfbottom{v3}
\fmfforce{(0.0625w,h)}{v1}
\fmfforce{(0.0625w,0)}{v3}
\fmffixed{(0.875w,0)}{v1,v2}
\fmffixed{(0.875w,0)}{v3,v4}
\fmffixed{(0,0.75h)}{vc3,vc2}
\fmf{plain}{v1,vc1}
\fmf{plain}{vc1,v3}
\fmf{plain}{v2,vc2}
\fmf{plain}{vc2,vc3}
\fmf{plain}{vc3,v4}
\fmffreeze
\fmf{photon,tension=0.5,left=0.125}{vc1,vc2}
\fmf{photon,tension=0.5,right=0.125}{vc1,vc3}
\end{fmfchar*}}}
{}+{}
\settoheight{\eqoff}{$\times$}%
\setlength{\eqoff}{0.5\eqoff}%
\addtolength{\eqoff}{-3.75\unitlength}%
\raisebox{\eqoff}{%
\fmfframe(0,0)(0,0){%
\begin{fmfchar*}(10,7.5)
\fmftop{v1}
\fmfbottom{v3}
\fmfforce{(0.0625w,h)}{v1}
\fmfforce{(0.0625w,0)}{v3}
\fmffixed{(0.875w,0)}{v1,v2}
\fmffixed{(0.875w,0)}{v3,v4}
\fmffixed{(0,0.75h)}{vc2,vc1}
\fmf{plain}{v1,vc1}
\fmf{plain}{vc1,vc2}
\fmf{plain}{vc2,v3}
\fmf{plain}{v2,vc3}
\fmf{plain}{vc3,v4}
\fmffreeze
\fmf{photon,tension=0.5,left=0.125}{vc1,vc3}
\fmf{photon,tension=0.5,right=0.125}{vc2,vc3}
\end{fmfchar*}}}
{}+{}
\settoheight{\eqoff}{$\times$}%
\setlength{\eqoff}{0.5\eqoff}%
\addtolength{\eqoff}{-3.75\unitlength}%
\raisebox{\eqoff}{%
\fmfframe(0,0)(0,0){%
\begin{fmfchar*}(10,7.5)
\fmftop{v1}
\fmfbottom{v3}
\fmfforce{(0.0625w,h)}{v1}
\fmfforce{(0.0625w,0)}{v3}
\fmffixed{(0.875w,0)}{v1,v2}
\fmffixed{(0.875w,0)}{v3,v4}
\fmffixed{(0,0.75h)}{vc3,vc2}
\fmf{plain}{v1,vc1}
\fmf{plain}{vc1,v3}
\fmf{plain}{v2,vc2}
\fmf{plain}{vc2,vc3}
\fmf{plain}{vc3,v4}
\fmffreeze
\fmf{photon,tension=0.5,left=0.125}{vc4,vc2}
\fmf{photon,tension=0.5,right=0.125}{vc4,vc3}
\fmf{photon}{vc1,vc4}
\end{fmfchar*}}}
{}+{}
\settoheight{\eqoff}{$\times$}%
\setlength{\eqoff}{0.5\eqoff}%
\addtolength{\eqoff}{-3.75\unitlength}%
\raisebox{\eqoff}{%
\fmfframe(0,0)(0,0){%
\begin{fmfchar*}(10,7.5)
\fmftop{v1}
\fmfbottom{v3}
\fmfforce{(0.0625w,h)}{v1}
\fmfforce{(0.0625w,0)}{v3}
\fmffixed{(0.875w,0)}{v1,v2}
\fmffixed{(0.875w,0)}{v3,v4}
\fmffixed{(0,0.75h)}{vc2,vc1}
\fmf{plain}{v1,vc1}
\fmf{plain}{vc1,vc2}
\fmf{plain}{vc2,v3}
\fmf{plain}{v2,vc3}
\fmf{plain}{vc3,v4}
\fmffreeze
\fmf{photon,tension=0.5,left=0.125}{vc1,vc4}
\fmf{photon,tension=0.5,right=0.125}{vc2,vc4}
\fmf{photon}{vc4,vc3}
\end{fmfchar*}}}
{}+{}
\settoheight{\eqoff}{$\times$}%
\setlength{\eqoff}{0.5\eqoff}%
\addtolength{\eqoff}{-3.75\unitlength}%
\raisebox{\eqoff}{%
\fmfframe(0,0)(0,0){%
\begin{fmfchar*}(10,7.5)
\fmftop{v1}
\fmfbottom{v3}
\fmfforce{(0.0625w,h)}{v1}
\fmfforce{(0.0625w,0)}{v3}
\fmffixed{(0.875w,0)}{v1,v2}
\fmffixed{(0.875w,0)}{v3,v4}
\fmffixed{(0,0.75h)}{vc2,vc1}
\fmffixed{(0,0.75h)}{vc4,vc3}
\fmf{plain}{v1,vc1}
\fmf{plain}{vc1,vc2}
\fmf{plain}{vc2,v3}
\fmf{plain}{v2,vc3}
\fmf{plain}{vc3,vc4}
\fmf{plain}{vc4,v4}
\fmffreeze
\fmf{photon,tension=0.5}{vc1,vc3}
\fmf{photon,tension=0.5}{vc2,vc4}
\end{fmfchar*}}}
{}+{}
\settoheight{\eqoff}{$\times$}%
\setlength{\eqoff}{0.5\eqoff}%
\addtolength{\eqoff}{-3.75\unitlength}%
\raisebox{\eqoff}{%
\fmfframe(0,0)(2,0){%
\begin{fmfchar*}(10,7.5)
\fmftop{v1}
\fmfbottom{v3}
\fmfforce{(0.0625w,h)}{v1}
\fmfforce{(0.0625w,0)}{v3}
\fmffixed{(0.875w,0)}{v1,v2}
\fmffixed{(0.875w,0)}{v3,v4}
\fmffixed{(0,0.75h)}{vc4,vc2}
\fmf{plain}{v1,vc1}
\fmf{plain}{vc1,v3}
\fmf{plain}{v2,vc2}
\fmf{plain}{vc2,vc3}
\fmf{plain}{vc3,vc4}
\fmf{plain}{vc4,v4}
\fmffreeze
\fmf{photon,tension=0.5,right=0.75}{vc4,vc2}
\fmf{photon,tension=0.5}{vc1,vc3}
\end{fmfchar*}}}
{}+{}
\settoheight{\eqoff}{$\times$}%
\setlength{\eqoff}{0.5\eqoff}%
\addtolength{\eqoff}{-3.75\unitlength}%
\raisebox{\eqoff}{%
\fmfframe(2,0)(0,0){%
\begin{fmfchar*}(10,7.5)
\fmftop{v1}
\fmfbottom{v3}
\fmfforce{(0.0625w,h)}{v1}
\fmfforce{(0.0625w,0)}{v3}
\fmffixed{(0.875w,0)}{v1,v2}
\fmffixed{(0.875w,0)}{v3,v4}
\fmffixed{(0,0.75h)}{vc3,vc1}
\fmf{plain}{v1,vc1}
\fmf{plain}{vc1,vc2}
\fmf{plain}{vc2,vc3}
\fmf{plain}{vc3,v3}
\fmf{plain}{v2,vc4}
\fmf{plain}{vc4,v4}
\fmffreeze
\fmf{photon,tension=0.5,right=0.75}{vc1,vc3}
\fmf{photon,tension=0.5}{vc4,vc2}
\end{fmfchar*}}}
\\
&\phantom{{}={}}
{}+{}
\settoheight{\eqoff}{$\times$}%
\setlength{\eqoff}{0.5\eqoff}%
\addtolength{\eqoff}{-3.75\unitlength}%
\raisebox{\eqoff}{%
\fmfframe(0,0)(0,0){%
\begin{fmfchar*}(10,7.5)
\fmftop{v1}
\fmfbottom{v3}
\fmfforce{(0.0625w,h)}{v1}
\fmfforce{(0.0625w,0)}{v3}
\fmffixed{(0.875w,0)}{v1,v2}
\fmffixed{(0.875w,0)}{v3,v4}
\fmf{plain}{v1,vc1}
\fmf{plain}{vc1,v3}
\fmf{plain}{v2,vc2}
\fmf{plain}{vc2,v4}
\fmffreeze
\fmf{photon}{vc1,vc3}
\fmf{photon}{vc4,vc2}
\fmf{plain,tension=0.33,left=0.5}{vc3,vc4}
\fmf{plain,tension=0.33,left=0.5}{vc4,vc3}
\end{fmfchar*}}}
{}+{}
\settoheight{\eqoff}{$\times$}%
\setlength{\eqoff}{0.5\eqoff}%
\addtolength{\eqoff}{-3.75\unitlength}%
\raisebox{\eqoff}{%
\fmfframe(0,0)(0,0){%
\begin{fmfchar*}(10,7.5)
\fmftop{v1}
\fmfbottom{v3}
\fmfforce{(0.0625w,h)}{v1}
\fmfforce{(0.0625w,0)}{v3}
\fmffixed{(0.875w,0)}{v1,v2}
\fmffixed{(0.875w,0)}{v3,v4}
\fmf{plain}{v1,vc1}
\fmf{plain}{vc1,v3}
\fmf{plain}{v2,vc2}
\fmf{plain}{vc2,v4}
\fmffreeze
\fmf{photon}{vc1,vc3}
\fmf{photon}{vc4,vc2}
\fmf{dashes,tension=0.33,left=0.5}{vc3,vc4}
\fmf{dashes,tension=0.33,left=0.5}{vc4,vc3}
\end{fmfchar*}}}
{}+{}
\settoheight{\eqoff}{$\times$}%
\setlength{\eqoff}{0.5\eqoff}%
\addtolength{\eqoff}{-3.75\unitlength}%
\raisebox{\eqoff}{%
\fmfframe(0,0)(0,0){%
\begin{fmfchar*}(10,7.5)
\fmftop{v1}
\fmfbottom{v3}
\fmfforce{(0.0625w,h)}{v1}
\fmfforce{(0.0625w,0)}{v3}
\fmffixed{(0.875w,0)}{v1,v2}
\fmffixed{(0.875w,0)}{v3,v4}
\fmf{plain}{v1,vc1}
\fmf{plain}{vc1,v3}
\fmf{plain}{v2,vc2}
\fmf{plain}{vc2,v4}
\fmffreeze
\fmf{photon}{vc1,vc3}
\fmf{photon}{vc4,vc2}
\fmf{photon,tension=0.33,left=0.5}{vc3,vc4}
\fmf{photon,tension=0.33,left=0.5}{vc4,vc3}
\end{fmfchar*}}}
{}+{}
\settoheight{\eqoff}{$\times$}%
\setlength{\eqoff}{0.5\eqoff}%
\addtolength{\eqoff}{-3.75\unitlength}%
\raisebox{\eqoff}{%
\fmfframe(0,0)(0,0){%
\begin{fmfchar*}(10,7.5)
\fmftop{v1}
\fmfbottom{v3}
\fmfforce{(0.0625w,h)}{v1}
\fmfforce{(0.0625w,0)}{v3}
\fmffixed{(0.875w,0)}{v1,v2}
\fmffixed{(0.875w,0)}{v3,v4}
\fmf{plain}{v1,vc1}
\fmf{plain}{vc1,v3}
\fmf{plain}{v2,vc2}
\fmf{plain}{vc2,v4}
\fmffreeze
\fmf{photon}{vc1,vc3}
\fmf{photon}{vc4,vc2}
\fmf{dots,tension=0.33,left=0.5}{vc3,vc4}
\fmf{dots,tension=0.33,left=0.5}{vc4,vc3}
\end{fmfchar*}}}
\pnt
\end{aligned}
\end{equation}
In the first term, which contains a fermion loop, the flavour 
flows from the upper to the respective lower external line as 
in all the other diagrams. 

The two-loop self-energy correction of the scalar field also appears 
as a sum of subdiagrams.
In terms of the non-vanishing diagrams it explicitly reads 
\begin{equation}\label{SigmaY}
\begin{aligned}
\Sigma_Y=
\settoheight{\eqoff}{$\times$}%
\setlength{\eqoff}{0.5\eqoff}%
\addtolength{\eqoff}{-3.75\unitlength}%
\raisebox{\eqoff}{%
\fmfframe(0,0)(0,0){%
\begin{fmfchar*}(10,7.5)
\fmfleft{v1}
\fmfright{v2}
\fmfforce{(0.0625w,0.5h)}{v1}
\fmfforce{(0.9375w,0.5h)}{v2}
\fmf{plain}{v1,v2}
\fmffreeze
\fmfposition
\vacpol{v1}{v2}
\end{fmfchar*}}}
&{}={}
\settoheight{\eqoff}{$\times$}%
\setlength{\eqoff}{0.5\eqoff}%
\addtolength{\eqoff}{-3.75\unitlength}%
\raisebox{\eqoff}{%
\fmfframe(0,0)(0,0){%
\begin{fmfchar*}(10,7.5)
\fmftop{v1}
\fmfbottom{v2}
\fmfforce{(0.0625w,0.5h)}{v1}
\fmfforce{(0.9375w,0.5h)}{v2}
\fmffixed{(0.65w,0)}{vc1,vc2}
\fmf{plain}{v1,vc1}
\fmf{plain}{vc1,vc2}
\fmf{plain}{vc2,v2}
\fmffreeze
\fmfposition
\fmf{dashes,left=0.875}{vc1,vc2}
\fmf{dashes,left=0.5}{vc1,vc2}
\end{fmfchar*}}}
{}+{}
\settoheight{\eqoff}{$\times$}%
\setlength{\eqoff}{0.5\eqoff}%
\addtolength{\eqoff}{-3.75\unitlength}%
\raisebox{\eqoff}{%
\fmfframe(0,0)(0,0){%
\begin{fmfchar*}(10,7.5)
\fmftop{v1}
\fmfbottom{v2}
\fmfforce{(0.0625w,0.5h)}{v1}
\fmfforce{(0.9375w,0.5h)}{v2}
\fmffixed{(0.65w,0)}{vc1,vc2}
\fmf{plain}{v1,vc1}
\fmf{plain}{vc1,vc2}
\fmf{plain}{vc2,v2}
\fmffreeze
\fmfposition
\fmf{dashes,right=0.875}{vc1,vc2}
\fmf{dashes,right=0.5}{vc1,vc2}
\end{fmfchar*}}}
{}+{}
\settoheight{\eqoff}{$\times$}%
\setlength{\eqoff}{0.5\eqoff}%
\addtolength{\eqoff}{-3.75\unitlength}%
\raisebox{\eqoff}{%
\fmfframe(0,0)(0,0){%
\begin{fmfchar*}(10,7.5)
\fmftop{v1}
\fmfbottom{v2}
\fmfforce{(0.0625w,0.5h)}{v1}
\fmfforce{(0.9375w,0.5h)}{v2}
\fmffixed{(0.65w,0)}{vc1,vc2}
\fmf{plain}{v1,vc1}
\fmf{plain}{vc1,vc2}
\fmf{plain}{vc2,v2}
\fmffreeze
\fmfposition
\fmf{dashes,left=0.875}{vc1,vc2}
\fmf{dashes,right=0.875}{vc1,vc2}
\end{fmfchar*}}}
{}+{}
\settoheight{\eqoff}{$\times$}%
\setlength{\eqoff}{0.5\eqoff}%
\addtolength{\eqoff}{-3.75\unitlength}%
\raisebox{\eqoff}{%
\fmfframe(0,0)(0,0){%
\begin{fmfchar*}(10,7.5)
\fmftop{v1}
\fmfbottom{v2}
\fmfforce{(0.0625w,0.5h)}{v1}
\fmfforce{(0.9375w,0.5h)}{v2}
\fmffixed{(0.65w,0)}{vc1,vc2}
\fmf{plain}{v1,vc1}
\fmf{plain}{vc1,vc2}
\fmf{plain}{vc2,v2}
\fmffreeze
\fmfposition
\fmf{photon,left=0.875}{vc1,vc2}
\fmf{photon,left=0.5}{vc1,vc2}
\end{fmfchar*}}}
{}+{}
\settoheight{\eqoff}{$\times$}%
\setlength{\eqoff}{0.5\eqoff}%
\addtolength{\eqoff}{-3.75\unitlength}%
\raisebox{\eqoff}{%
\fmfframe(0,0)(0,0){%
\begin{fmfchar*}(10,7.5)
\fmftop{v1}
\fmfbottom{v2}
\fmfforce{(0.0625w,0.5h)}{v1}
\fmfforce{(0.9375w,0.5h)}{v2}
\fmffixed{(0.65w,0)}{vc1,vc2}
\fmf{plain}{v1,vc1}
\fmf{plain}{vc1,vc2}
\fmf{plain}{vc2,v2}
\fmffreeze
\fmfposition
\fmf{photon,right=0.875}{vc1,vc2}
\fmf{photon,right=0.5}{vc1,vc2}
\end{fmfchar*}}}
{}+{}
\settoheight{\eqoff}{$\times$}%
\setlength{\eqoff}{0.5\eqoff}%
\addtolength{\eqoff}{-3.75\unitlength}%
\raisebox{\eqoff}{%
\fmfframe(0,0)(0,0){%
\begin{fmfchar*}(10,7.5)
\fmftop{v1}
\fmfbottom{v2}
\fmfforce{(0.0625w,0.5h)}{v1}
\fmfforce{(0.9375w,0.5h)}{v2}
\fmffixed{(0.65w,0)}{vc1,vc2}
\fmf{plain}{v1,vc1}
\fmf{plain}{vc1,vc2}
\fmf{plain}{vc2,v2}
\fmffreeze
\fmfposition
\fmf{photon,left=0.875}{vc1,vc2}
\fmf{photon,right=0.875}{vc1,vc2}
\end{fmfchar*}}}
\\
&\phantom{{}={}}
{}+{}
\settoheight{\eqoff}{$\times$}%
\setlength{\eqoff}{0.5\eqoff}%
\addtolength{\eqoff}{-3.75\unitlength}%
\raisebox{\eqoff}{%
\fmfframe(0,0)(0,0){%
\begin{fmfchar*}(10,7.5)
\fmftop{v1}
\fmfbottom{v2}
\fmfforce{(0.0625w,0.5h)}{v1}
\fmfforce{(0.9375w,0.5h)}{v2}
\fmffixed{(0.65w,0)}{vc1,vc3}
\fmffixed{(0.325w,0)}{vc1,vc2}
\fmf{plain}{v1,vc1}
\fmf{plain}{vc1,vc2}
\fmf{plain}{vc2,vc3}
\fmf{plain}{vc3,v2}
\fmffreeze
\fmfposition
\fmf{photon,left=1}{vc1,vc2}
\fmf{photon,left=1}{vc2,vc3}
\end{fmfchar*}}}
{}+{}
\settoheight{\eqoff}{$\times$}%
\setlength{\eqoff}{0.5\eqoff}%
\addtolength{\eqoff}{-3.75\unitlength}%
\raisebox{\eqoff}{%
\fmfframe(0,0)(0,0){%
\begin{fmfchar*}(10,7.5)
\fmftop{v1}
\fmfbottom{v2}
\fmfforce{(0.0625w,0.5h)}{v1}
\fmfforce{(0.9375w,0.5h)}{v2}
\fmffixed{(0.65w,0)}{vc1,vc3}
\fmffixed{(0.325w,0)}{vc1,vc2}
\fmf{plain}{v1,vc1}
\fmf{plain}{vc1,vc2}
\fmf{plain}{vc2,vc3}
\fmf{plain}{vc3,v2}
\fmffreeze
\fmfposition
\fmf{photon,right=1}{vc1,vc2}
\fmf{photon,right=1}{vc2,vc3}
\end{fmfchar*}}}
{}+{}
\settoheight{\eqoff}{$\times$}%
\setlength{\eqoff}{0.5\eqoff}%
\addtolength{\eqoff}{-3.75\unitlength}%
\raisebox{\eqoff}{%
\fmfframe(0,0)(0,0){%
\begin{fmfchar*}(10,7.5)
\fmftop{v1}
\fmfbottom{v2}
\fmfforce{(0.0625w,0.5h)}{v1}
\fmfforce{(0.9375w,0.5h)}{v2}
\fmffixed{(0.65w,0)}{vc1,vc3}
\fmffixed{(0.325w,0)}{vc1,vc2}
\fmf{plain}{v1,vc1}
\fmf{plain}{vc1,vc2}
\fmf{plain}{vc2,vc3}
\fmf{plain}{vc3,v2}
\fmffreeze
\fmfposition
\fmf{photon,left=1}{vc1,vc2}
\fmf{photon,right=1}{vc2,vc3}
\end{fmfchar*}}}
{}+{}
\settoheight{\eqoff}{$\times$}%
\setlength{\eqoff}{0.5\eqoff}%
\addtolength{\eqoff}{-3.75\unitlength}%
\raisebox{\eqoff}{%
\fmfframe(0,0)(0,0){%
\begin{fmfchar*}(10,7.5)
\fmftop{v1}
\fmfbottom{v2}
\fmfforce{(0.0625w,0.5h)}{v1}
\fmfforce{(0.9375w,0.5h)}{v2}
\fmffixed{(0.65w,0)}{vc1,vc3}
\fmffixed{(0.325w,0)}{vc1,vc2}
\fmf{plain}{v1,vc1}
\fmf{plain}{vc1,vc2}
\fmf{plain}{vc2,vc3}
\fmf{plain}{vc3,v2}
\fmffreeze
\fmfposition
\fmf{photon,right=1}{vc1,vc2}
\fmf{photon,left=1}{vc2,vc3}
\end{fmfchar*}}}
{}+{}
\settoheight{\eqoff}{$\times$}%
\setlength{\eqoff}{0.5\eqoff}%
\addtolength{\eqoff}{-3.75\unitlength}%
\raisebox{\eqoff}{%
\fmfframe(0,0)(0,0){%
\begin{fmfchar*}(10,7.5)
\fmftop{v1}
\fmfbottom{v2}
\fmfforce{(0.0625w,0.5h)}{v1}
\fmfforce{(0.9375w,0.5h)}{v2}
\fmffixed{(0.2w,0)}{vc1,vc2}
\fmffixed{(0.2w,0)}{vc2,vc3}
\fmffixed{(0.2w,0)}{vc3,vc4}
\fmf{plain}{v1,vc1}
\fmf{plain}{vc1,vc2}
\fmf{plain}{vc2,vc3}
\fmf{plain}{vc3,vc4}
\fmf{plain}{vc4,v2}
\fmffreeze
\fmfposition
\fmf{photon,left=1}{vc1,vc3}
\fmf{photon,right=1}{vc2,vc4}
\end{fmfchar*}}}
{}+{}
\settoheight{\eqoff}{$\times$}%
\setlength{\eqoff}{0.5\eqoff}%
\addtolength{\eqoff}{-3.75\unitlength}%
\raisebox{\eqoff}{%
\fmfframe(0,0)(0,0){%
\begin{fmfchar*}(10,7.5)
\fmftop{v1}
\fmfbottom{v2}
\fmfforce{(0.0625w,0.5h)}{v1}
\fmfforce{(0.9375w,0.5h)}{v2}
\fmffixed{(0.2w,0)}{vc1,vc2}
\fmffixed{(0.2w,0)}{vc2,vc3}
\fmffixed{(0.2w,0)}{vc3,vc4}
\fmf{plain}{v1,vc1}
\fmf{plain}{vc1,vc2}
\fmf{plain}{vc2,vc3}
\fmf{plain}{vc3,vc4}
\fmf{plain}{vc4,v2}
\fmffreeze
\fmfposition
\fmf{photon,right=1}{vc1,vc3}
\fmf{photon,left=1}{vc2,vc4}
\end{fmfchar*}}}
{}+{}
\settoheight{\eqoff}{$\times$}%
\setlength{\eqoff}{0.5\eqoff}%
\addtolength{\eqoff}{-3.75\unitlength}%
\raisebox{\eqoff}{%
\fmfframe(0,0)(0,0){%
\begin{fmfchar*}(10,7.5)
\fmftop{v1}
\fmfbottom{v2}
\fmfforce{(0.0625w,0.5h)}{v1}
\fmfforce{(0.9375w,0.5h)}{v2}
\fmffixed{(0.65w,0)}{vc1,vc3}
\fmffixed{(0.325w,0)}{vc1,vc2}
\fmfforce{(0.5w,h)}{vc4}
\fmf{plain}{v1,vc1}
\fmf{plain}{vc1,vc2}
\fmf{plain}{vc2,vc3}
\fmf{plain}{vc3,v2}
\fmffreeze
\fmfposition
\fmf{photon,left=0.5}{vc1,vc4}
\fmf{photon}{vc4,vc2}
\fmf{photon,left=0.5}{vc4,vc3}
\end{fmfchar*}}}
{}+{}
\settoheight{\eqoff}{$\times$}%
\setlength{\eqoff}{0.5\eqoff}%
\addtolength{\eqoff}{-3.75\unitlength}%
\raisebox{\eqoff}{%
\fmfframe(0,0)(0,0){%
\begin{fmfchar*}(10,7.5)
\fmftop{v1}
\fmfbottom{v2}
\fmfforce{(0.0625w,0.5h)}{v1}
\fmfforce{(0.9375w,0.5h)}{v2}
\fmffixed{(0.65w,0)}{vc1,vc3}
\fmffixed{(0.325w,0)}{vc1,vc2}
\fmfforce{(0.5w,0)}{vc4}
\fmf{plain}{v1,vc1}
\fmf{plain}{vc1,vc2}
\fmf{plain}{vc2,vc3}
\fmf{plain}{vc3,v2}
\fmffreeze
\fmfposition
\fmf{photon,right=0.5}{vc1,vc4}
\fmf{photon}{vc4,vc2}
\fmf{photon,right=0.5}{vc4,vc3}
\end{fmfchar*}}}
\\
&\phantom{{}={}}
{}+{}
\settoheight{\eqoff}{$\times$}%
\setlength{\eqoff}{0.5\eqoff}%
\addtolength{\eqoff}{-3.75\unitlength}%
\raisebox{\eqoff}{%
\fmfframe(0,0)(0,0){%
\begin{fmfchar*}(10,7.5)
\fmftop{v1}
\fmfbottom{v2}
\fmfforce{(0.0625w,0.5h)}{v1}
\fmfforce{(0.9375w,0.5h)}{v2}
\fmffixed{(0.65w,0)}{vc1,vc2}
\fmffixed{(whatever,0.4h)}{vc1,vc3}
\fmffixed{(0.4w,0)}{vc3,vc4}
\fmf{plain}{v1,vc1}
\fmf{plain}{vc1,vc2}
\fmf{plain}{vc2,v2}
\fmf{photon,left=0.5}{vc1,vc3}
\fmf{plain,left=0.5}{vc3,vc4}
\fmf{plain,right=0.5}{vc3,vc4}
\fmf{photon,right=0.5}{vc2,vc4}
\end{fmfchar*}}}
{}+{}
\settoheight{\eqoff}{$\times$}%
\setlength{\eqoff}{0.5\eqoff}%
\addtolength{\eqoff}{-3.75\unitlength}%
\raisebox{\eqoff}{%
\fmfframe(0,0)(0,0){%
\begin{fmfchar*}(10,7.5)
\fmftop{v1}
\fmfbottom{v2}
\fmfforce{(0.0625w,0.5h)}{v1}
\fmfforce{(0.9375w,0.5h)}{v2}
\fmffixed{(0.65w,0)}{vc1,vc2}
\fmffixed{(whatever,-0.4h)}{vc1,vc3}
\fmffixed{(0.4w,0)}{vc3,vc4}
\fmf{plain}{v1,vc1}
\fmf{plain}{vc1,vc2}
\fmf{plain}{vc2,v2}
\fmf{photon,right=0.5}{vc1,vc3}
\fmf{plain,left=0.5}{vc3,vc4}
\fmf{plain,right=0.5}{vc3,vc4}
\fmf{photon,left=0.5}{vc2,vc4}
\end{fmfchar*}}}
{}+{}
\settoheight{\eqoff}{$\times$}%
\setlength{\eqoff}{0.5\eqoff}%
\addtolength{\eqoff}{-3.75\unitlength}%
\raisebox{\eqoff}{%
\fmfframe(0,0)(0,0){%
\begin{fmfchar*}(10,7.5)
\fmftop{v1}
\fmfbottom{v2}
\fmfforce{(0.0625w,0.5h)}{v1}
\fmfforce{(0.9375w,0.5h)}{v2}
\fmffixed{(0.65w,0)}{vc1,vc2}
\fmffixed{(whatever,0.4h)}{vc1,vc3}
\fmffixed{(0.4w,0)}{vc3,vc4}
\fmf{plain}{v1,vc1}
\fmf{plain}{vc1,vc2}
\fmf{plain}{vc2,v2}
\fmf{photon,left=0.5}{vc1,vc3}
\fmf{dashes,left=0.5}{vc3,vc4}
\fmf{dashes,right=0.5}{vc3,vc4}
\fmf{photon,right=0.5}{vc2,vc4}
\end{fmfchar*}}}
{}+{}
\settoheight{\eqoff}{$\times$}%
\setlength{\eqoff}{0.5\eqoff}%
\addtolength{\eqoff}{-3.75\unitlength}%
\raisebox{\eqoff}{%
\fmfframe(0,0)(0,0){%
\begin{fmfchar*}(10,7.5)
\fmftop{v1}
\fmfbottom{v2}
\fmfforce{(0.0625w,0.5h)}{v1}
\fmfforce{(0.9375w,0.5h)}{v2}
\fmffixed{(0.65w,0)}{vc1,vc2}
\fmffixed{(whatever,-0.4h)}{vc1,vc3}
\fmffixed{(0.4w,0)}{vc3,vc4}
\fmf{plain}{v1,vc1}
\fmf{plain}{vc1,vc2}
\fmf{plain}{vc2,v2}
\fmf{photon,right=0.5}{vc1,vc3}
\fmf{dashes,left=0.5}{vc3,vc4}
\fmf{dashes,right=0.5}{vc3,vc4}
\fmf{photon,left=0.5}{vc2,vc4}
\end{fmfchar*}}}
{}+{}
\settoheight{\eqoff}{$\times$}%
\setlength{\eqoff}{0.5\eqoff}%
\addtolength{\eqoff}{-3.75\unitlength}%
\raisebox{\eqoff}{%
\fmfframe(0,0)(0,0){%
\begin{fmfchar*}(10,7.5)
\fmftop{v1}
\fmfbottom{v2}
\fmfforce{(0.0625w,0.5h)}{v1}
\fmfforce{(0.9375w,0.5h)}{v2}
\fmffixed{(0.65w,0)}{vc1,vc2}
\fmffixed{(whatever,0.4h)}{vc1,vc3}
\fmffixed{(0.4w,0)}{vc3,vc4}
\fmf{plain}{v1,vc1}
\fmf{plain}{vc1,vc2}
\fmf{plain}{vc2,v2}
\fmf{photon,left=0.5}{vc1,vc3}
\fmf{photon,left=0.5}{vc3,vc4}
\fmf{photon,right=0.5}{vc3,vc4}
\fmf{photon,right=0.5}{vc2,vc4}
\end{fmfchar*}}}
{}+{}
\settoheight{\eqoff}{$\times$}%
\setlength{\eqoff}{0.5\eqoff}%
\addtolength{\eqoff}{-3.75\unitlength}%
\raisebox{\eqoff}{%
\fmfframe(0,0)(0,0){%
\begin{fmfchar*}(10,7.5)
\fmftop{v1}
\fmfbottom{v2}
\fmfforce{(0.0625w,0.5h)}{v1}
\fmfforce{(0.9375w,0.5h)}{v2}
\fmffixed{(0.65w,0)}{vc1,vc2}
\fmffixed{(whatever,-0.4h)}{vc1,vc3}
\fmffixed{(0.4w,0)}{vc3,vc4}
\fmf{plain}{v1,vc1}
\fmf{plain}{vc1,vc2}
\fmf{plain}{vc2,v2}
\fmf{photon,right=0.5}{vc1,vc3}
\fmf{photon,left=0.5}{vc3,vc4}
\fmf{photon,right=0.5}{vc3,vc4}
\fmf{photon,left=0.5}{vc2,vc4}
\end{fmfchar*}}}
{}+{}
\settoheight{\eqoff}{$\times$}%
\setlength{\eqoff}{0.5\eqoff}%
\addtolength{\eqoff}{-3.75\unitlength}%
\raisebox{\eqoff}{%
\fmfframe(0,0)(0,0){%
\begin{fmfchar*}(10,7.5)
\fmftop{v1}
\fmfbottom{v2}
\fmfforce{(0.0625w,0.5h)}{v1}
\fmfforce{(0.9375w,0.5h)}{v2}
\fmffixed{(0.65w,0)}{vc1,vc2}
\fmffixed{(whatever,0.4h)}{vc1,vc3}
\fmffixed{(0.4w,0)}{vc3,vc4}
\fmf{plain}{v1,vc1}
\fmf{plain}{vc1,vc2}
\fmf{plain}{vc2,v2}
\fmf{photon,left=0.5}{vc1,vc3}
\fmf{dots,left=0.5}{vc3,vc4}
\fmf{dots,right=0.5}{vc3,vc4}
\fmf{photon,right=0.5}{vc2,vc4}
\end{fmfchar*}}}
{}+{}
\settoheight{\eqoff}{$\times$}%
\setlength{\eqoff}{0.5\eqoff}%
\addtolength{\eqoff}{-3.75\unitlength}%
\raisebox{\eqoff}{%
\fmfframe(0,0)(0,0){%
\begin{fmfchar*}(10,7.5)
\fmftop{v1}
\fmfbottom{v2}
\fmfforce{(0.0625w,0.5h)}{v1}
\fmfforce{(0.9375w,0.5h)}{v2}
\fmffixed{(0.65w,0)}{vc1,vc2}
\fmffixed{(whatever,-0.4h)}{vc1,vc3}
\fmffixed{(0.4w,0)}{vc3,vc4}
\fmf{plain}{v1,vc1}
\fmf{plain}{vc1,vc2}
\fmf{plain}{vc2,v2}
\fmf{photon,right=0.5}{vc1,vc3}
\fmf{dots,left=0.5}{vc3,vc4}
\fmf{dots,right=0.5}{vc3,vc4}
\fmf{photon,left=0.5}{vc2,vc4}
\end{fmfchar*}}}
\pnt
\end{aligned}
\end{equation} 
To calculate the divergences of the corresponding four-loop diagrams, 
we need the divergent 
and finite parts of the two-loop self-energy.
Apart from the Wick rotation (the result has to be multiplied by a factor 
$i^2=-1$), the amputated scalar self-energy contribution becomes 
\begin{equation}\label{sY}
\begin{aligned}
\Sigma_Y
=
\settoheight{\eqoff}{$\times$}%
\setlength{\eqoff}{0.5\eqoff}%
\addtolength{\eqoff}{-3.75\unitlength}%
\raisebox{\eqoff}{%
\fmfframe(0,0)(0,0){%
\begin{fmfchar*}(10,7.5)
\fmfleft{v1}
\fmfright{v2}
\fmfforce{(0.0625w,0.5h)}{v1}
\fmfforce{(0.9375w,0.5h)}{v2}
\fmf{plain}{v1,v2}
\fmffreeze
\fmfposition
\vacpol{v1}{v2}
\end{fmfchar*}}}
&=i
\Big[\frac{\lambda\hat\lambda}{4}\Big(\frac{3}{2\varepsilon}-\frac{3}{2}\pi^2
+3\Big(\frac{25}{3}-\gamma+\ln4\pi\Big)\Big)\\
&\phantom{{}=i\Big[}
+\frac{(\lambda-\hat\lambda)^2}{4}\Big(\frac{1}{4\varepsilon}-\frac{\pi^2}{4}+\frac{1}{2}(3-\gamma+\ln4\pi)\Big)\Big]
\settoheight{\eqoff}{$\times$}%
\setlength{\eqoff}{0.5\eqoff}%
\addtolength{\eqoff}{-3.75\unitlength}%
\raisebox{\eqoff}{%
\fmfframe(1,0)(1,0){%
\begin{fmfchar*}(10,7.5)
\fmfleft{v1}
\fmfright{v2}
\fmfforce{(0.0625w,0.5h)}{v1}
\fmfforce{(0.9375w,0.5h)}{v2}
\fmf{plain,label=$\scriptscriptstyle -1+2\varepsilon$,l.side=left,l.dist=2}{v1,v2}
\fmffreeze
\fmfposition
\end{fmfchar*}}}
\col
\end{aligned}
\end{equation}
where the last propagator factor on the r.h.s.\ captures the momentum dependence. Its weight label indicates the exponent of $\frac{1}{p^2}$, where $p$
is the external momentum. The pole part of the above result coincides with 
the result in \cite{Bak:2008vd}.

The amputated fermionic one-loop self energy is given by
\begin{equation}\label{Sigmapsi}
\begin{aligned}
\Sigma_\psi=
\settoheight{\eqoff}{$\times$}%
\setlength{\eqoff}{0.5\eqoff}%
\addtolength{\eqoff}{-3.75\unitlength}%
\raisebox{\eqoff}{%
\fmfframe(0,0)(0,0){%
\begin{fmfchar*}(10,7.5)
\fmfleft{v1}
\fmfright{v2}
\fmfforce{(0.0625w,0.5h)}{v1}
\fmfforce{(0.9375w,0.5h)}{v2}
\fmf{dashes}{v1,v2}
\fmffreeze
\fmfposition
\vacpol{v1}{v2}
\end{fmfchar*}}}
&=
\settoheight{\eqoff}{$\times$}%
\setlength{\eqoff}{0.5\eqoff}%
\addtolength{\eqoff}{-3.75\unitlength}%
\raisebox{\eqoff}{%
\fmfframe(0,0)(0,0){%
\begin{fmfchar*}(10,7.5)
\fmfleft{v1}
\fmfright{v2}
\fmfforce{(0.0625w,0.5h)}{v1}
\fmfforce{(0.9375w,0.5h)}{v2}
\fmffixed{(0.65w,0)}{vc1,vc2}
\fmf{phantom}{v1,vc1}
\fmf{phantom}{vc1,vc2}
\fmf{phantom}{vc2,v2}
\fmf{photon,left=0.875}{vc1,vc2}
\fmf{dashes}{v1,v2}
\fmffreeze
\fmfposition
\end{fmfchar*}}}
+
\settoheight{\eqoff}{$\times$}%
\setlength{\eqoff}{0.5\eqoff}%
\addtolength{\eqoff}{-3.75\unitlength}%
\raisebox{\eqoff}{%
\fmfframe(0,0)(0,0){%
\begin{fmfchar*}(10,7.5)
\fmfleft{v1}
\fmfright{v2}
\fmfforce{(0.0625w,0.5h)}{v1}
\fmfforce{(0.9375w,0.5h)}{v2}
\fmffixed{(0.65w,0)}{vc1,vc2}
\fmf{phantom}{v1,vc1}
\fmf{phantom}{vc1,vc2}
\fmf{phantom}{vc2,v2}
\fmf{photon,right=0.875}{vc1,vc2}
\fmf{dashes}{v1,v2}
\fmffreeze
\fmfposition
\end{fmfchar*}}}
=-i(\lambda-\hat\lambda)\frac{1}{16}
\settoheight{\eqoff}{$\times$}%
\setlength{\eqoff}{0.5\eqoff}%
\addtolength{\eqoff}{-3.75\unitlength}%
\raisebox{\eqoff}{%
\fmfframe(1,0)(1,0){%
\begin{fmfchar*}(10,7.5)
\fmfleft{v1}
\fmfright{v2}
\fmfforce{(0.0625w,0.5h)}{v1}
\fmfforce{(0.9375w,0.5h)}{v2}
\fmf{plain,label=$\scriptscriptstyle -\frac{1}{2}+\varepsilon$,l.side=left,l.dist=2}{v1,v2}
\fmffreeze
\fmfposition
\end{fmfchar*}}}
\col
\end{aligned}
\end{equation}
where the left and right fermion fields respectively are $\psi$ and $\psi^\dagger$.

\subsection{Diagrams involving only the scalar six-vertex}

The diagrams which lead to non-trivial flavour permutations are given by 
\begin{equation}\label{Sgraphs}
\begin{aligned}
S_2&=
\settoheight{\eqoff}{$\times$}%
\setlength{\eqoff}{0.5\eqoff}%
\addtolength{\eqoff}{-8.5\unitlength}%
\raisebox{\eqoff}{%
\fmfframe(0,1)(5,1){%
\begin{fmfchar*}(20,15)
\fmftop{v5}
\fmfbottom{v6}
\fmfforce{(0.125w,h)}{v5}
\fmfforce{(0.125w,0)}{v6}
\fmffixed{(0.25w,0)}{v2,v1}
\fmffixed{(0.25w,0)}{v3,v2}
\fmffixed{(0.25w,0)}{v4,v3}
\fmffixed{(0.25w,0)}{v5,v4}
\fmffixed{(0.25w,0)}{v6,v7}
\fmffixed{(0.25w,0)}{v7,v8}
\fmffixed{(0.25w,0)}{v8,v9}
\fmffixed{(0.25w,0)}{v9,v10}
\fmffixed{(whatever,0)}{v61,v62}
\fmf{plain,left=0.25}{v1,v62}
\fmf{plain}{v2,v62}
\fmf{phantom,right=0.25}{v3,v62}
\fmf{plain,tension=0.5,left=0.25}{v8,v62}
\fmf{plain,tension=0.5}{v9,v62}
\fmf{plain,tension=0.5,right=0.25}{v10,v62}
\fmf{plain,left=0.25}{v6,v61}
\fmf{plain}{v7,v61}
\fmf{phantom,right=0.25}{v8,v61}
\fmf{plain,tension=0.5,left=0.25}{v3,v61}
\fmf{plain,tension=0.5}{v4,v61}
\fmf{plain,tension=0.5,right=0.25}{v5,v61}
\fmffreeze
\fmf{plain}{v61,v62}
\fmf{plain,tension=1,left=0,width=1mm}{v6,v10}
\end{fmfchar*}}}
\to 
\frac{(\lambda\hat\lambda)^2}{16}\Big(-\frac{1}{2\varepsilon^2}+\frac{2}{\varepsilon}\Big)(\ptwo{1}{3}-\pone{1})
\col\\
S_4&=
\settoheight{\eqoff}{$\times$}%
\setlength{\eqoff}{0.5\eqoff}%
\addtolength{\eqoff}{-8.5\unitlength}%
\raisebox{\eqoff}{%
\fmfframe(0,1)(0,1){%
\begin{fmfchar*}(20,15)
\fmftop{v4}
\fmfbottom{v5}
\fmfforce{(0.125w,h)}{v4}
\fmfforce{(0.125w,0)}{v5}
\fmffixed{(0.25w,0)}{v2,v1}
\fmffixed{(0.25w,0)}{v3,v2}
\fmffixed{(0.25w,0)}{v4,v3}
\fmffixed{(0.25w,0)}{v5,v6}
\fmffixed{(0.25w,0)}{v6,v7}
\fmffixed{(0.25w,0)}{v7,v8}
\fmffixed{(whatever,0)}{v61,v62}
\fmf{plain,left=0.25}{v1,v62}
\fmf{phantom,right=0.25}{v3,v62}
\fmf{plain,tension=0.5,left=0.25}{v6,v62}
\fmf{plain,tension=0.5}{v7,v62}
\fmf{plain,tension=0.5,right=0.25}{v8,v62}
\fmf{plain,left=0.25}{v5,v61}
\fmf{phantom,right=0.25}{v7,v61}
\fmf{plain,tension=0.5,left=0.25}{v2,v61}
\fmf{plain,tension=0.5}{v3,v61}
\fmf{plain,tension=0.5,right=0.25}{v4,v61}
\fmffreeze
\fmf{plain,left=0.5}{v61,v62}
\fmf{plain,left=0.5}{v62,v61}
\fmf{plain,tension=1,left=0,width=1mm}{v5,v8}
\end{fmfchar*}}}
\to
\frac{(\lambda\hat\lambda)^2}{16}\Big(-\frac{1}{2\varepsilon^2}+\frac{2}{\varepsilon}\Big)
(\ptwo{1}{2}-\pone{1})
\col\\
S_5&=
\settoheight{\eqoff}{$\times$}%
\setlength{\eqoff}{0.5\eqoff}%
\addtolength{\eqoff}{-8.5\unitlength}%
\raisebox{\eqoff}{%
\fmfframe(0,1)(0,1){%
\begin{fmfchar*}(20,15)
\fmftop{v4}
\fmfbottom{v5}
\fmfforce{(0.125w,h)}{v4}
\fmfforce{(0.125w,0)}{v5}
\fmffixed{(0.25w,0)}{v2,v1}
\fmffixed{(0.25w,0)}{v3,v2}
\fmffixed{(0.25w,0)}{v4,v3}
\fmffixed{(0.25w,0)}{v5,v6}
\fmffixed{(0.25w,0)}{v6,v7}
\fmffixed{(0.25w,0)}{v7,v8}
\fmffixed{(whatever,0)}{v61,v62}
\fmf{plain,left=0.25}{v1,v62}
\fmf{plain,right=0.25}{v2,v62}
\fmf{plain,tension=0.5,left=0.25}{v7,v62}
\fmf{plain,tension=0.5,right=0.25}{v8,v62}
\fmf{plain,left=0.25}{v5,v61}
\fmf{plain,right=0.25}{v6,v61}
\fmf{plain,tension=0.5,left=0.25}{v3,v61}
\fmf{plain,tension=0.5,right=0.25}{v4,v61}
\fmffreeze
\fmf{plain,left=0.25}{v61,v62}
\fmf{plain,left=0.25}{v62,v61}
\fmf{plain,tension=1,left=0,width=1mm}{v5,v8}
\end{fmfchar*}}}
\to
\frac{\lambda^3\hat\lambda}{16}\frac{\pi^2}{8\varepsilon}(\pone{1}+\pone{2})
\col\\
S_6&=
\settoheight{\eqoff}{$\times$}%
\setlength{\eqoff}{0.5\eqoff}%
\addtolength{\eqoff}{-8.5\unitlength}%
\raisebox{\eqoff}{%
\fmfframe(0,1)(-5,1){%
\begin{fmfchar*}(20,15)
\fmftop{v3}
\fmfbottom{v4}
\fmfforce{(0.125w,h)}{v3}
\fmfforce{(0.125w,0)}{v4}
\fmffixed{(0.25w,0)}{v2,v1}
\fmffixed{(0.25w,0)}{v3,v2}
\fmffixed{(0.25w,0)}{v4,v5}
\fmffixed{(0.25w,0)}{v5,v6}
\fmffixed{(0,0.33h)}{v61,v62}
\fmf{plain,tension=0.5,left=0.25}{v1,v62}
\fmf{plain,tension=0.5}{v2,v62}
\fmf{plain,tension=0.5,right=0.25}{v3,v62}
\fmf{plain,tension=0.5,left=0.25}{v4,v61}
\fmf{plain,tension=0.5}{v5,v61}
\fmf{plain,tension=0.5,right=0.25}{v6,v61}
\fmffreeze
\fmf{plain,left=0.5}{v61,v62}
\fmf{plain}{v62,v61}
\fmf{plain,right=0.5}{v61,v62}
\fmf{plain,tension=1,left=0,width=1mm}{v4,v6}
\end{fmfchar*}}}
\to
\frac{(\lambda\hat\lambda)^2}{16}\frac{1}{\varepsilon^2}\pone{1}
\col\\
S_7&=
\settoheight{\eqoff}{$\times$}%
\setlength{\eqoff}{0.5\eqoff}%
\addtolength{\eqoff}{-8.5\unitlength}%
\raisebox{\eqoff}{%
\fmfframe(0,1)(-5,1){%
\begin{fmfchar*}(20,15)
\fmftop{v3}
\fmfbottom{v4}
\fmfforce{(0.125w,h)}{v3}
\fmfforce{(0.125w,0)}{v4}
\fmffixed{(0.25w,0)}{v2,v1}
\fmffixed{(0.25w,0)}{v3,v2}
\fmffixed{(0.25w,0)}{v4,v5}
\fmffixed{(0.25w,0)}{v5,v6}
\fmffixed{(whatever,0)}{v61,v62}
\fmf{plain,left=0.25}{v1,v62}
\fmf{phantom,right=0.25}{v2,v62}
\fmf{plain,tension=0.5,left=0.25}{v5,v62}
\fmf{plain,tension=0.5,right=0.25}{v6,v62}
\fmf{plain,left=0.25}{v4,v61}
\fmf{phantom,right=0.25}{v5,v61}
\fmf{plain,tension=0.5,left=0.25}{v2,v61}
\fmf{plain,tension=0.5,right=0.25}{v3,v61}
\fmffreeze
\fmf{plain,left=0.5}{v61,v62}
\fmf{plain}{v62,v61}
\fmf{plain,right=0.5}{v61,v62}
\fmf{plain,tension=1,left=0,width=1mm}{v4,v6}
\end{fmfchar*}}}
\to
\frac{(\lambda\hat\lambda)^2}{16}\Big(\frac{1}{2\varepsilon^2}-\frac{2}{\varepsilon}\Big)\pone{1}
\pnt
\end{aligned}
\end{equation}
As explained at the beginning of this section, we have
neglected all contributions to the trace and identity operator.  
We have also not evaluated the remaining scalar  diagrams
\begin{equation}
\begin{aligned}
S_1=
\settoheight{\eqoff}{$\times$}%
\setlength{\eqoff}{0.5\eqoff}%
\addtolength{\eqoff}{-8.5\unitlength}%
\raisebox{\eqoff}{%
\fmfframe(0,1)(5,1){%
\begin{fmfchar*}(20,15)
\fmftop{v5}
\fmfbottom{v6}
\fmfforce{(0.125w,h)}{v5}
\fmfforce{(0.125w,0)}{v6}
\fmffixed{(0.25w,0)}{v2,v1}
\fmffixed{(0.25w,0)}{v3,v2}
\fmffixed{(0.25w,0)}{v4,v3}
\fmffixed{(0.25w,0)}{v5,v4}
\fmffixed{(0.25w,0)}{v6,v7}
\fmffixed{(0.25w,0)}{v7,v8}
\fmffixed{(0.25w,0)}{v8,v9}
\fmffixed{(0.25w,0)}{v9,v10}
\fmffixed{(whatever,0)}{v61,v62}
\fmf{plain,left=0.25}{v1,v62}
\fmf{phantom,right=0.25}{v4,v62}
\fmf{plain,tension=0.5,left=0.25}{v7,v62}
\fmf{plain,tension=0.5,left=0.125}{v8,v62}
\fmf{plain,tension=0.5,right=0.125}{v9,v62}
\fmf{plain,tension=0.5,right=0.25}{v10,v62}
\fmf{plain,left=0.25}{v6,v61}
\fmf{phantom,right=0.25}{v9,v61}
\fmf{plain,tension=0.5,left=0.25}{v2,v61}
\fmf{plain,tension=0.5,left=0.125}{v3,v61}
\fmf{plain,tension=0.5,right=0.125}{v4,v61}
\fmf{plain,tension=0.5,right=0.25}{v5,v61}
\fmffreeze
\fmf{plain}{v61,v62}
\fmf{plain,tension=1,left=0,width=1mm}{v6,v10}
\end{fmfchar*}}}
\col\quad
S_3=
\settoheight{\eqoff}{$\times$}%
\setlength{\eqoff}{0.5\eqoff}%
\addtolength{\eqoff}{-8.5\unitlength}%
\raisebox{\eqoff}{%
\fmfframe(0,1)(0,1){%
\begin{fmfchar*}(20,15)
\fmftop{v4}
\fmfbottom{v5}
\fmfforce{(0.125w,h)}{v4}
\fmfforce{(0.125w,0)}{v5}
\fmffixed{(0.25w,0)}{v2,v1}
\fmffixed{(0.25w,0)}{v3,v2}
\fmffixed{(0.25w,0)}{v4,v3}
\fmffixed{(0.25w,0)}{v5,v6}
\fmffixed{(0.25w,0)}{v6,v7}
\fmffixed{(0.25w,0)}{v7,v8}
\fmffixed{(0,0.33h)}{v61,v62}
\fmf{plain,tension=0.5,left=0.25}{v1,v62}
\fmf{plain,tension=0.5,left=0.125}{v2,v62}
\fmf{plain,tension=0.5,right=0.125}{v3,v62}
\fmf{plain,tension=0.5,right=0.25}{v4,v62}
\fmf{plain,tension=0.5,left=0.25}{v5,v61}
\fmf{plain,tension=0.5,left=0.125}{v6,v61}
\fmf{plain,tension=0.5,right=0.125}{v7,v61}
\fmf{plain,tension=0.5,right=0.25}{v8,v61}
\fmffreeze
\fmf{plain,left=0.5}{v61,v62}
\fmf{plain,left=0.5}{v62,v61}
\fmf{plain,tension=1,left=0,width=1mm}{v5,v8}
\end{fmfchar*}}}
\col\quad
S_8=
\settoheight{\eqoff}{$\times$}%
\setlength{\eqoff}{0.5\eqoff}%
\addtolength{\eqoff}{-8.5\unitlength}%
\raisebox{\eqoff}{%
\fmfframe(0,1)(-10,1){%
\begin{fmfchar*}(20,15)
\fmftop{v2}
\fmfbottom{v3}
\fmfforce{(0.125w,h)}{v2}
\fmfforce{(0.125w,0)}{v3}
\fmffixed{(0.25w,0)}{v2,v1}
\fmffixed{(0.25w,0)}{v3,v4}
\fmffixed{(0,0.33h)}{v61,v62}
\fmf{plain,tension=0.5,left=0.25}{v1,v62}
\fmf{plain,tension=0.5,right=0.25}{v2,v62}
\fmf{plain,tension=0.5,left=0.25}{v3,v61}
\fmf{plain,tension=0.5,right=0.25}{v4,v61}
\fmffreeze
\fmf{plain,left=0.75}{v61,v62}
\fmf{plain,right=0.25}{v62,v61}
\fmf{plain,right=0.25}{v61,v62}
\fmf{plain,left=0.75}{v62,v61}
\fmf{plain,tension=1,left=0,width=1mm}{v3,v4}
\end{fmfchar*}}}
\col\quad
S_9=
\settoheight{\eqoff}{$\times$}%
\setlength{\eqoff}{0.5\eqoff}%
\addtolength{\eqoff}{-8.5\unitlength}%
\raisebox{\eqoff}{%
\fmfframe(0,1)(-10,1){%
\begin{fmfchar*}(20,15)
\fmftop{v2}
\fmfbottom{v3}
\fmfforce{(0.125w,h)}{v2}
\fmfforce{(0.125w,0)}{v3}
\fmffixed{(0.25w,0)}{v2,v1}
\fmffixed{(0.25w,0)}{v3,v4}
\fmffixed{(whatever,0)}{v61,v62}
\fmf{plain}{v1,v62}
\fmf{plain}{v4,v62}
\fmf{plain}{v2,v61}
\fmf{plain}{v3,v61}
\fmffreeze
\fmf{plain,left=0.75}{v61,v62}
\fmf{plain,right=0.25}{v62,v61}
\fmf{plain,right=0.25}{v61,v62}
\fmf{plain,left=0.75}{v62,v61}
\fmf{plain,tension=1,left=0,width=1mm}{v3,v4}
\end{fmfchar*}}}
\col\quad
S_{10}=
\settoheight{\eqoff}{$\times$}%
\setlength{\eqoff}{0.5\eqoff}%
\addtolength{\eqoff}{-8.5\unitlength}%
\raisebox{\eqoff}{%
\fmfframe(-6,1)(-6,1){%
\begin{fmfchar*}(20,15)
\fmftop{v1}
\fmfbottom{v2}
\fmfforce{(0.5w,h)}{v1}
\fmfforce{(0.5w,0)}{v2}
\fmffixed{(0.001w,0)}{v2i,v2}
\fmffixed{(0.001w,0)}{v2,v2o}
\fmffixed{(0,whatever)}{v61,v62}
\fmf{plain}{v1,v62}
\fmf{plain}{v2,v61}
\fmffixed{(0,0.33h)}{v61,v62}
\fmffreeze
\fmf{plain,left=1}{v61,v62}
\fmf{plain,right=0.5}{v62,v61}
\fmf{plain}{v62,v61}
\fmf{plain,right=0.5}{v61,v62}
\fmf{plain,left=1}{v62,v61}
\fmf{plain,tension=1,left=0,width=1mm}{v2i,v2o}
\end{fmfchar*}}}
\end{aligned}
\end{equation}
which only contribute to the trace or identity operator.

Along with the set of diagrams in \eqref{Sgraphs}, we also need
the reflected diagrams of $S_2$, $S_4$, $S_7$, 
which are easily obtained by
using the transformations in appendix \ref{app:symmetries}. 
From the diagram $S_5$  and from its 
shifted version, in which $\la$ and $\hla$ are interchanged, we have to 
consider the terms involving the permutation $\pone{1}$.
The sum of the relevant set of diagrams with two scalar six-vertices
then reads
\begin{equation}\label{S}
\begin{aligned}
S
&=
\frac{\lambda\hat\lambda}{16}
\Big[\lambda\hat\lambda\Big(\Big(-\frac{1}{2\varepsilon^2}+\frac{2}{\varepsilon}\Big)(\ptwo{1}{3}+\ptwo{3}{1}+2\ptwo{1}{2})
+\Big(\frac{4}{\varepsilon^2}-\frac{1}{\varepsilon}\Big(12-\frac{1}{4}\pi^2\Big)\Big)\pone{1}\Big)
\\
&\phantom{{}\to{}\frac{\lambda\hat\lambda}{16}\Big[}
+(\lambda-\hat\lambda)^2\frac{\pi^2}{8\varepsilon}\pone{1}
\Big]
\pnt
\end{aligned}
\end{equation}

\subsection{Diagrams involving a fermion square}

The only four-loop diagrams in which four neighboured scalar lines interact, and which contain a fermion square loop are given by
\begin{equation}
\begin{aligned}
F_{\mathbf{s}1}(\lambda\hat\lambda)
&=
\settoheight{\eqoff}{$\times$}%
\setlength{\eqoff}{0.5\eqoff}%
\addtolength{\eqoff}{-11\unitlength}%
\raisebox{\eqoff}{%
\fmfframe(0,1)(0,1){%
\begin{fmfchar*}(20,20)
\fmftop{v1}
\fmfbottom{v5}
\fmfforce{(0.125w,h)}{v1}
\fmfforce{(0.125w,0)}{v5}
\fmffixed{(0.25w,0)}{v1,v2}
\fmffixed{(0.25w,0)}{v2,v3}
\fmffixed{(0.25w,0)}{v3,v4}
\fmffixed{(0.25w,0)}{v5,v6}
\fmffixed{(0.25w,0)}{v6,v7}
\fmffixed{(0.25w,0)}{v7,v8}
\fmffixed{(0,whatever)}{vc2,vc4}
\fmffixed{(0.5w,0)}{vc1,vc3}
\fmf{plain,tension=1,right=0.125}{v1,vc1}
\fmf{plain,tension=0.25,right=0.25}{v2,vc2}
\fmf{plain,tension=0.25,left=0.25}{v3,vc2}
\fmf{plain,tension=1,left=0.125}{v4,vc3}
\fmf{plain,tension=1,left=0.125}{v5,vc1}
\fmf{plain,tension=0.25,left=0.25}{v6,vc4}
\fmf{plain,tension=0.25,right=0.25}{v7,vc4}
\fmf{plain,tension=1,right=0.125}{v8,vc3}
 \fmf{dashes,tension=0.5,left=0.25}{vc1,vc2}
\fmf{dashes,tension=0.5,left=0.25}{vc2,vc3}
\fmf{dashes,tension=0.5,left=0.25}{vc4,vc1}
\fmf{dashes,tension=0.5,left=0.25}{vc3,vc4}
\fmffreeze
\fmfposition
\fmf{plain,tension=1,left=0,width=1mm}{v5,v8}
\fmffreeze
\end{fmfchar*}}}
=-\frac{(\lambda\hat\lambda)^2}{16}\frac{4}{\varepsilon}
(\ptwo{1}{2}-\pone{1})
\col\qquad
&F_{\mathbf{s}2}(\lambda\hat\lambda)
&=
\settoheight{\eqoff}{$\times$}%
\setlength{\eqoff}{0.5\eqoff}%
\addtolength{\eqoff}{-11\unitlength}%
\raisebox{\eqoff}{%
\fmfframe(0,1)(0,1){%
\begin{fmfchar*}(20,20)
\fmftop{v1}
\fmfbottom{v5}
\fmfforce{(0.125w,h)}{v1}
\fmfforce{(0.125w,0)}{v5}
\fmffixed{(0.25w,0)}{v1,v2}
\fmffixed{(0.25w,0)}{v2,v3}
\fmffixed{(0.25w,0)}{v3,v4}
\fmffixed{(0.25w,0)}{v5,v6}
\fmffixed{(0.25w,0)}{v6,v7}
\fmffixed{(0.25w,0)}{v7,v8}
%
\fmf{plain,tension=0.5,right=0.25}{v1,vc1}
\fmf{plain,tension=0.5,left=0.25}{v2,vc1}
\fmf{plain,tension=0.5,right=0.25}{v3,vc2}
\fmf{plain,tension=0.5,left=0.25}{v4,vc2}
  \fmf{dashes}{vc1,vc3}
  \fmf{dashes}{vc2,vc4}
\fmf{plain,tension=0.5,left=0.25}{v5,vc3}
\fmf{plain,tension=0.5,right=0.25}{v6,vc3}
\fmf{plain,tension=0.5,left=0.25}{v7,vc4}
\fmf{plain,tension=0.5,right=0.25}{v8,vc4}
\fmf{plain,tension=0.5,right=0,width=1mm}{v5,v8}
\fmffreeze
  \fmf{dashes}{vc1,vc2}
  \fmf{dashes}{vc4,vc3}
\end{fmfchar*}}}
\col
\end{aligned}
\end{equation}
where the second diagram only contributes to the identity and trace part 
of the dilatation operator and hence is not considered here.

Besides the above diagrams, there are the following diagrams in which only 
three neighbouring lines interact. They contain either a bubble and 
are given by
\begin{equation}
\begin{aligned}
F_{\mathbf{sb}1}(\lambda\hat\lambda)
&=
\settoheight{\eqoff}{$\times$}%
\setlength{\eqoff}{0.5\eqoff}%
\addtolength{\eqoff}{-11\unitlength}%
\raisebox{\eqoff}{%
\fmfframe(0,1)(-5,1){%
\begin{fmfchar*}(20,20)
\fmftop{v1}
\fmfbottom{v5}
\fmfforce{(0.125w,h)}{v1}
\fmfforce{(0.125w,0)}{v5}
\fmffixed{(0.25w,0)}{v1,v2}
\fmffixed{(0.25w,0)}{v2,v3}
\fmffixed{(0.25w,0)}{v3,v4}
\fmffixed{(0.25w,0)}{v5,v6}
\fmffixed{(0.25w,0)}{v6,v7}
\fmffixed{(0.25w,0)}{v7,v8}
%
\fmf{plain,tension=0.5,right=0.25}{v1,vc1}
\fmf{plain,tension=0.5,left=0.25}{v2,vc1}
\fmf{plain,tension=0.5,left=0.25}{v5,vc3}
\fmf{plain,tension=0.5,right=0.25}{v6,vc3}
  \fmf{dashes}{vc1,vc3}
\fmffreeze
\fmfpoly{phantom}{vc2,vc1,vc3,vc4}
\fmf{plain,left=0.125}{v3,vc2}
\fmf{plain,right=0.125}{v7,vc4}
  \fmf{plain,right=0.5}{vc2,vc4}
  \fmf{dashes,left=0.5}{vc2,vc4}
  \fmf{dashes}{vc1,vc2}
  \fmf{dashes}{vc4,vc3}
\fmf{plain,tension=0.5,right=0,width=1mm}{v5,v7}
\end{fmfchar*}}}
\to-\frac{(\lambda\hat\lambda)^2}{16}\frac{1}{\varepsilon^2}\pone{1}
\col\qquad
&F_{\mathbf{sb}2}(\lambda\hat\lambda)
&=
\settoheight{\eqoff}{$\times$}%
\setlength{\eqoff}{0.5\eqoff}%
\addtolength{\eqoff}{-11\unitlength}%
\raisebox{\eqoff}{%
\fmfframe(0,1)(-5,1){%
\begin{fmfchar*}(20,20)
\fmftop{v1}
\fmfbottom{v5}
\fmfforce{(0.125w,h)}{v1}
\fmfforce{(0.125w,0)}{v5}
\fmffixed{(0.25w,0)}{v1,v2}
\fmffixed{(0.25w,0)}{v2,v3}
\fmffixed{(0.25w,0)}{v3,v4}
\fmffixed{(0.25w,0)}{v5,v6}
\fmffixed{(0.25w,0)}{v6,v7}
\fmffixed{(0.25w,0)}{v7,v8}
%
\fmf{plain,tension=0.5,right=0.25}{v1,vc1}
\fmf{plain,tension=0.5,left=0.25}{v2,vc1}
\fmf{plain,tension=0.5,left=0.25}{v5,vc3}
\fmf{plain,tension=0.5,right=0.25}{v6,vc3}
  \fmf{dashes}{vc1,vc3}
\fmffreeze
\fmfpoly{phantom}{vc2,vc1,vc3,vc4}
\fmf{plain,left=0.125}{v3,vc2}
\fmf{plain,right=0.125}{v7,vc4}
  \fmf{dashes,right=0.5}{vc2,vc4}
  \fmf{plain,left=0.5}{vc2,vc4}
  \fmf{dashes}{vc1,vc2}
  \fmf{dashes}{vc4,vc3}
\fmf{plain,tension=0.5,right=0,width=1mm}{v5,v7}
\end{fmfchar*}}}
\to-\frac{(\lambda\hat\lambda)^2}{16}\frac{1}{2\varepsilon^2}\pone{1}
\col\\
F_{\mathbf{sb}3}(\lambda\hat\lambda)
&=
\settoheight{\eqoff}{$\times$}%
\setlength{\eqoff}{0.5\eqoff}%
\addtolength{\eqoff}{-11\unitlength}%
\raisebox{\eqoff}{%
\fmfframe(0,1)(-5,1){%
\begin{fmfchar*}(20,20)
\fmftop{v1}
\fmfbottom{v5}
\fmfforce{(0.125w,h)}{v1}
\fmfforce{(0.125w,0)}{v5}
\fmffixed{(0.25w,0)}{v1,v2}
\fmffixed{(0.25w,0)}{v2,v3}
\fmffixed{(0.25w,0)}{v3,v4}
\fmffixed{(0.25w,0)}{v5,v6}
\fmffixed{(0.25w,0)}{v6,v7}
\fmffixed{(0.25w,0)}{v7,v8}
%
\fmf{plain}{v2,vc1}
\fmf{plain}{v6,vc4}
  \fmf{phantom}{vc1,vc4}
\fmffreeze
\fmf{plain,tension=0.5,left=0.25}{v3,vc1}
\fmf{plain,tension=0.5,right=0.125}{v7,vc2}
\fmfpoly{phantom}{vc2,vc1,vc3,vc4}
\fmf{plain,right=0.125}{v1,vc3}
\fmf{plain,left=0.125}{v5,vc3}
  \fmf{plain,right=0.5}{vc2,vc4}
  \fmf{dashes,left=0.5}{vc2,vc4}
  \fmf{dashes}{vc1,vc3}
  \fmf{dashes}{vc1,vc2}
  \fmf{dashes}{vc4,vc3}
\fmf{plain,tension=0.5,right=0,width=1mm}{v5,v7}
\end{fmfchar*}}}
\to-\frac{(\lambda\hat\lambda)^2}{16}\frac{1}{\varepsilon^2}\pone{1}
\col\qquad
&F_{\mathbf{sb}4}(\lambda\hat\lambda)
&=
\settoheight{\eqoff}{$\times$}%
\setlength{\eqoff}{0.5\eqoff}%
\addtolength{\eqoff}{-11\unitlength}%
\raisebox{\eqoff}{%
\fmfframe(0,1)(-5,1){%
\begin{fmfchar*}(20,20)
\fmftop{v1}
\fmfbottom{v5}
\fmfforce{(0.125w,h)}{v1}
\fmfforce{(0.125w,0)}{v5}
\fmffixed{(0.25w,0)}{v1,v2}
\fmffixed{(0.25w,0)}{v2,v3}
\fmffixed{(0.25w,0)}{v3,v4}
\fmffixed{(0.25w,0)}{v5,v6}
\fmffixed{(0.25w,0)}{v6,v7}
\fmffixed{(0.25w,0)}{v7,v8}
%
\fmf{plain}{v2,vc1}
\fmf{plain}{v6,vc4}
  \fmf{phantom}{vc1,vc4}
\fmffreeze
\fmf{plain,tension=0.5,left=0.25}{v3,vc1}
\fmf{plain,tension=0.5,right=0.125}{v7,vc2}
\fmfpoly{phantom}{vc2,vc1,vc3,vc4}
\fmf{plain,right=0.125}{v1,vc3}
\fmf{plain,left=0.125}{v5,vc3}
  \fmf{dashes,right=0.5}{vc2,vc4}
  \fmf{plain,left=0.5}{vc2,vc4}
  \fmf{dashes}{vc1,vc3}
  \fmf{dashes}{vc1,vc2}
  \fmf{dashes}{vc4,vc3}
\fmf{plain,tension=0.5,right=0,width=1mm}{v5,v7}
\end{fmfchar*}}}
\to-\frac{(\lambda\hat\lambda)^2}{16}\frac{1}{2\varepsilon^2}\pone{1}
\col\\
F_{\mathbf{sb}5}(\lambda\hat\lambda)
&=
\settoheight{\eqoff}{$\times$}%
\setlength{\eqoff}{0.5\eqoff}%
\addtolength{\eqoff}{-11\unitlength}%
\raisebox{\eqoff}{%
\fmfframe(0,1)(-5,1){%
\begin{fmfchar*}(20,20)
\fmftop{v1}
\fmfbottom{v5}
\fmfforce{(0.125w,h)}{v1}
\fmfforce{(0.125w,0)}{v5}
\fmffixed{(0.25w,0)}{v1,v2}
\fmffixed{(0.25w,0)}{v2,v3}
\fmffixed{(0.25w,0)}{v3,v4}
\fmffixed{(0.25w,0)}{v5,v6}
\fmffixed{(0.25w,0)}{v6,v7}
\fmffixed{(0.25w,0)}{v7,v8}
%
\fmf{plain}{v6,vc1}
\fmf{plain}{v2,vc4}
  \fmf{phantom}{vc1,vc4}
\fmffreeze
\fmf{plain,tension=0.5,right=0.25}{v7,vc1}
\fmf{plain,tension=0.5,left=0.125}{v3,vc3}
\fmfpoly{phantom}{vc2,vc1,vc3,vc4}
\fmf{plain,left=0.125}{v5,vc2}
\fmf{plain,right=0.125}{v1,vc2}
  \fmf{plain,left=0.5}{vc3,vc4}
  \fmf{dashes,right=0.5}{vc3,vc4}
  \fmf{dashes}{vc1,vc2}
  \fmf{dashes}{vc1,vc3}
  \fmf{dashes}{vc4,vc2}
\fmf{plain,tension=0.5,right=0,width=1mm}{v5,v7}
\end{fmfchar*}}}
\to-\frac{(\lambda\hat\lambda)^2}{16}\frac{1}{\varepsilon^2}\pone{1}
\col\qquad
&F_{\mathbf{sb}6}(\lambda\hat\lambda)
&=
\settoheight{\eqoff}{$\times$}%
\setlength{\eqoff}{0.5\eqoff}%
\addtolength{\eqoff}{-11\unitlength}%
\raisebox{\eqoff}{%
\fmfframe(0,1)(-5,1){%
\begin{fmfchar*}(20,20)
\fmftop{v1}
\fmfbottom{v5}
\fmfforce{(0.125w,h)}{v1}
\fmfforce{(0.125w,0)}{v5}
\fmffixed{(0.25w,0)}{v1,v2}
\fmffixed{(0.25w,0)}{v2,v3}
\fmffixed{(0.25w,0)}{v3,v4}
\fmffixed{(0.25w,0)}{v5,v6}
\fmffixed{(0.25w,0)}{v6,v7}
\fmffixed{(0.25w,0)}{v7,v8}
%
\fmf{plain}{v6,vc1}
\fmf{plain}{v2,vc4}
  \fmf{phantom}{vc1,vc4}
\fmffreeze
\fmf{plain,tension=0.5,right=0.25}{v7,vc1}
\fmf{plain,tension=0.5,left=0.125}{v3,vc3}
\fmfpoly{phantom}{vc2,vc1,vc3,vc4}
\fmf{plain,left=0.125}{v5,vc2}
\fmf{plain,right=0.125}{v1,vc2}
  \fmf{dashes,left=0.5}{vc3,vc4}
  \fmf{plain,right=0.5}{vc3,vc4}
  \fmf{dashes}{vc1,vc2}
  \fmf{dashes}{vc1,vc3}
  \fmf{dashes}{vc4,vc2}
\fmf{plain,tension=0.5,right=0,width=1mm}{v5,v7}
\end{fmfchar*}}}
\to-\frac{(\lambda\hat\lambda)^2}{16}\frac{1}{2\varepsilon^2}\pone{1}
\col
\end{aligned}
\end{equation}
or they contain triangles and read
\begin{equation}
\begin{aligned}
F_{\mathbf{st}1}(\lambda\hat\lambda)
&=
\settoheight{\eqoff}{$\times$}%
\setlength{\eqoff}{0.5\eqoff}%
\addtolength{\eqoff}{-11\unitlength}%
\raisebox{\eqoff}{%
\fmfframe(0,1)(-5,1){%
\begin{fmfchar*}(20,20)
\fmftop{v1}
\fmfbottom{v5}
\fmfforce{(0.125w,h)}{v1}
\fmfforce{(0.125w,0)}{v5}
\fmffixed{(0.25w,0)}{v1,v2}
\fmffixed{(0.25w,0)}{v2,v3}
\fmffixed{(0.25w,0)}{v3,v4}
\fmffixed{(0.25w,0)}{v5,v6}
\fmffixed{(0.25w,0)}{v6,v7}
\fmffixed{(0.25w,0)}{v7,v8}
%
\fmf{plain}{v6,vc1}
\fmf{plain}{v2,vc4}
  \fmf{plain}{vc1,vc4}
\fmffreeze
\fmf{plain,tension=0.5,right=0.125}{v7,vc3}
\fmf{plain,tension=0.5,left=0.125}{v3,vc3}
\fmfpoly{phantom}{vc2,vc1,vc3,vc4}
\fmf{plain,left=0.125}{v5,vc2}
\fmf{plain,right=0.125}{v1,vc2}
  \fmf{dashes}{vc3,vc4}
  \fmf{dashes}{vc1,vc2}
  \fmf{dashes}{vc1,vc3}
  \fmf{dashes}{vc4,vc2}
\fmf{plain,tension=0.5,right=0,width=1mm}{v5,v7}
\end{fmfchar*}}}
\to\frac{(\lambda\hat\lambda)^2}{16}\Big(\frac{1}{\varepsilon^2}
-\frac{1}{\varepsilon}\Big(4-\frac{2}{3}\pi^2\Big)\Big)\pone{1}
\col\\
F_{\mathbf{st}2}(\lambda\hat\lambda)
&=
\settoheight{\eqoff}{$\times$}%
\setlength{\eqoff}{0.5\eqoff}%
\addtolength{\eqoff}{-11\unitlength}%
\raisebox{\eqoff}{%
\fmfframe(0,1)(-5,1){%
\begin{fmfchar*}(20,20)
\fmftop{v1}
\fmfbottom{v5}
\fmfforce{(0.125w,h)}{v1}
\fmfforce{(0.125w,0)}{v5}
\fmffixed{(0.25w,0)}{v1,v2}
\fmffixed{(0.25w,0)}{v2,v3}
\fmffixed{(0.25w,0)}{v3,v4}
\fmffixed{(0.25w,0)}{v5,v6}
\fmffixed{(0.25w,0)}{v6,v7}
\fmffixed{(0.25w,0)}{v7,v8}
\fmffixed{(0.25w,0)}{vc1,vc2}
\fmf{plain,right=0.125}{v1,vc1}
\fmf{phantom,right=0.125}{v2,vc1}
\fmf{plain,right=0.125}{v2,vc2}
\fmf{plain,left=0.125}{v3,vc2}
\fmf{plain,left=0.125}{v5,vc3}
\fmf{plain,right=0.125}{v6,vc3}
\fmf{phantom,right=0.125}{v6,vc4}
\fmf{plain,right=0.125}{v7,vc4}
\fmfpoly{phantom}{vc2,vc1,vc3,vc4}
\fmffreeze
  \fmf{dashes}{vc3,vc4}
  \fmf{dashes}{vc1,vc2}
  \fmf{dashes}{vc1,vc3}
  \fmf{dashes}{vc4,vc2}
  \fmf{plain}{vc1,vc4}
\fmf{plain,tension=0.5,right=0,width=1mm}{v5,v7}
\end{fmfchar*}}}
\to\frac{(\lambda\hat\lambda)^2}{16}\frac{1}{\varepsilon^2}\pone{1}
\pnt\\
\end{aligned}
\end{equation}
Summing up the above contributions, 
allowing for  factors
of two 
that come from reflections of the diagrams 
$F_{\mathbf{sb}1}(\lambda\hat\lambda)$ to 
$F_{\mathbf{sb}6}(\lambda\hat\lambda)$ and 
$F_{\mathbf{st}2}(\lambda\hat\lambda)$, we obtain
\begin{equation}\label{Fs}
\begin{aligned}
F_{\mathbf{s}}
&=\frac{(\lambda\hat\lambda)^2}{16}\Big[
-\frac{4}{\varepsilon}\ptwo{1}{2}
+\Big(-\frac{6}{\varepsilon^2}
+\frac{2}{3\varepsilon}\pi^2\Big)\pone{1}
\Big]\pnt
\end{aligned}
\end{equation}

\subsection{Diagrams involving fermion triangles}

The four-loop diagrams in which three quartic scalar fermion vertices form a
fermion triangle also involve a single gluon propagator.
The relevant diagrams are given by
\begin{equation}\label{Ftgraphs}
\begin{aligned}
F_{\mathbf{ts}1}(\lambda,\hat\lambda)
&=
\settoheight{\eqoff}{$\times$}%
\setlength{\eqoff}{0.5\eqoff}%
\addtolength{\eqoff}{-8.5\unitlength}%
\smash[b]{%
\raisebox{\eqoff}{%
\fmfframe(0,1)(-5,1){%
\begin{fmfchar*}(20,15)
\ftrianglerangethree
\vacpol{vt1}{vt2}
\fmf{plain,tension=1,left=0,width=1mm}{v4,v6}
\end{fmfchar*}}}}
=
\settoheight{\eqoff}{$\times$}%
\setlength{\eqoff}{0.5\eqoff}%
\addtolength{\eqoff}{-8.5\unitlength}%
\raisebox{\eqoff}{%
\fmfframe(0,1)(-5,1){%
\begin{fmfchar*}(20,15)
\ftrianglerangethree
\vacpol{vt3}{vt1}
\fmf{plain,tension=1,left=0,width=1mm}{v4,v6}
\end{fmfchar*}}}
\to
z(\lambda-\hat\lambda)\lambda^2\hat\lambda\frac{1}{16}\Big(-\frac{1}{2\varepsilon^2}\Big)\pone{1}
\col\\
F_{\mathbf{ts}2}(\lambda,\hat\lambda)
&=
\settoheight{\eqoff}{$\times$}%
\setlength{\eqoff}{0.5\eqoff}%
\addtolength{\eqoff}{-8.5\unitlength}%
\raisebox{\eqoff}{%
\fmfframe(0,1)(-5,1){%
\begin{fmfchar*}(20,15)
\ftrianglerangethree
\vacpol{vt3}{vt2}
\fmf{plain,tension=1,left=0,width=1mm}{v4,v6}
\end{fmfchar*}}}
\to
z(\lambda-\hat\lambda)\lambda^2\hat\lambda\frac{1}{16}\Big(-\frac{1}{2\varepsilon^2}
+\frac{2}{\varepsilon}\Big)\pone{1}
\col\\
F_{\mathbf{tv}1}(\lambda,\hat\lambda)
&=
\settoheight{\eqoff}{$\times$}%
\setlength{\eqoff}{0.5\eqoff}%
\addtolength{\eqoff}{-8.5\unitlength}%
\smash[b]{%
\raisebox{\eqoff}{%
\fmfframe(0,1)(-5,1){%
\begin{fmfchar*}(20,15)
\ftrianglerangethree
\fmf{photon}{vc1,vc2}
\fmf{plain,tension=1,left=0,width=1mm}{v4,v6}
\end{fmfchar*}}}}
=
\settoheight{\eqoff}{$\times$}%
\setlength{\eqoff}{0.5\eqoff}%
\addtolength{\eqoff}{-8.5\unitlength}%
\raisebox{\eqoff}{%
\fmfframe(0,1)(-5,1){%
\begin{fmfchar*}(20,15)
\ftrianglerangethree
\fmf{photon}{vc2,vc3}
\fmf{plain,tension=1,left=0,width=1mm}{v4,v6}
\end{fmfchar*}}}
\to
z\frac{\lambda^3\hat\lambda}{16}\Big(\frac{1}{2\varepsilon^2}
-\frac{1}{\varepsilon}\Big(1-\frac{\pi^2}{4}\Big)\Big)\pone{1}
\col\\
F_{\mathbf{tv}2}(\lambda,\hat\lambda)
&=
\settoheight{\eqoff}{$\times$}%
\setlength{\eqoff}{0.5\eqoff}%
\addtolength{\eqoff}{-8.5\unitlength}%
\raisebox{\eqoff}{%
\fmfframe(0,1)(-5,1){%
\begin{fmfchar*}(20,15)
\ftrianglerangethree
\fmf{photon}{vc3,vc1}
\fmf{plain,tension=1,left=0,width=1mm}{v4,v6}
\end{fmfchar*}}}
\to
z\frac{\lambda^3\hat\lambda}{16}\frac{1}{2\varepsilon^2}\pone{1}
\col\\
F_{\mathbf{tv}3}(\lambda\hat\lambda)
&=
\settoheight{\eqoff}{$\times$}%
\setlength{\eqoff}{0.5\eqoff}%
\addtolength{\eqoff}{-8.5\unitlength}%
\raisebox{\eqoff}{%
\fmfframe(0,1)(-5,1){%
\begin{fmfchar*}(20,15)
\ftrianglerangethree
\fmfi{photon}{vloc(__vc3){dir 180}..vm4}
\fmf{plain,tension=1,left=0,width=1mm}{v4,v6}
\end{fmfchar*}}}
\to
z\frac{(\lambda\hat\lambda)^2}{16}\frac{1}{\varepsilon}
\Big(6-\frac{2}{3}\pi^2\Big)\pone{1}
\col\\
F_{\mathbf{tv}4}(\lambda\hat\lambda)
&=
\settoheight{\eqoff}{$\times$}%
\setlength{\eqoff}{0.5\eqoff}%
\addtolength{\eqoff}{-8.5\unitlength}%
\raisebox{\eqoff}{%
\fmfframe(0,1)(-5,1){%
\begin{fmfchar*}(20,15)
\ftrianglerangethree
\fmfi{photon}{vloc(__vc2){dir -60}..vm5}
\fmf{plain,tension=1,left=0,width=1mm}{v4,v6}
\end{fmfchar*}}}
\to
z\frac{(\lambda\hat\lambda)^2}{16}\frac{1}{\varepsilon}
\Big(3-\frac{\pi^2}{4}\Big)\pone{1}
\col\\
F_{\mathbf{tv}5}(\lambda\hat\lambda)
&=
\settoheight{\eqoff}{$\times$}%
\setlength{\eqoff}{0.5\eqoff}%
\addtolength{\eqoff}{-8.5\unitlength}%
\raisebox{\eqoff}{%
\fmfframe(0,1)(-5,1){%
\begin{fmfchar*}(20,15)
\ftrianglerangethree
\fmfi{photon}{vloc(__vc2){dir -60}..vm6}
\fmf{plain,tension=1,left=0,width=1mm}{v4,v6}
\end{fmfchar*}}}
\to
z\frac{(\lambda\hat\lambda)^2}{16}\frac{1}{\varepsilon}
\Big(-1+\frac{\pi^2}{4}\Big)\pone{1}
\pnt
\end{aligned}
\end{equation}
The result depends on a sign $z=\pm1$ which is not uniquely given in
the literature \cite{Bak:2008cp,Benna:2008zy}. 
We will fix $z$ by computing the 
renormalization of the scalar six vertex at two loops. It should
vanish for the correct sign choice of $z$ to ensure superconformal 
invariance. 
 
Considering also the reflected diagrams, constructed by using the 
transformation rules of appendix \ref{app:symmetries}, 
we obtain for the sum of the above diagrams
\begin{equation}
\begin{aligned}\label{Ft}
F_{\mathbf{t}}
&=
z\frac{\lambda\hat\lambda}{16}\Big[
\lambda\hat\lambda\Big(
\frac{3}{\varepsilon^2}+\frac{1}{\varepsilon}\Big(12-\frac{\pi^2}{3}\Big)\Big)
+(\lambda-\hat\lambda)^2\frac{\pi^2}{2\varepsilon}\Big]\pone{1}
\pnt
\end{aligned}
\end{equation}
The presence of the reflected diagrams which do not differ by a
sign from the original diagram, but only by an exchange 
$\lambda\leftrightarrow\hat\lambda$ thereby guarantees that the result
only depends quadratically on the difference of the couplings and
hence on the parameter $\sigma$ defined in \eqref{barlambdasigmadef}.

\subsection{Diagrams involving a single scalar six-vertex and flavour-neutral substructures}

The two-loop diagram involving a single scalar six-vertex is promoted to  
logarithmically divergent four-loop diagrams by adding appropriate 
flavour-neutral substructures. 
We find diagrams which are completed by the next-to-nearest or nearest neighbour interactions given in \eqref{nnnint} and \eqref{nnint} or by the two-loop 
self energy correction of the scalar field \eqref{SigmaY}. 
Keeping only the contributions to the non-trivial permutation, we find
\begin{equation}
\begin{aligned}
S_{\mathbf{n}1}
&=
\settoheight{\eqoff}{$\times$}%
\setlength{\eqoff}{0.5\eqoff}%
\addtolength{\eqoff}{-8.5\unitlength}%
\raisebox{\eqoff}{%
\fmfframe(0,1)(0,1){%
\begin{fmfchar*}(20,15)
\vsixrangefourl
\nnint{vm6}{vem6}
\fmf{plain,tension=1,left=0,width=1mm}{v4,ved1}
\end{fmfchar*}}}
\to\frac{\lambda^3\hat\lambda}{16}
\Big(-\frac{1}{8\varepsilon^2}-\frac{3}{8\varepsilon}\pi^2\Big)\pone{1}
\col\\
S_{\mathbf{n}2}
&=
\settoheight{\eqoff}{$\times$}%
\setlength{\eqoff}{0.5\eqoff}%
\addtolength{\eqoff}{-8.5\unitlength}%
\raisebox{\eqoff}{%
\fmfframe(0,1)(0,1){%
\begin{fmfchar*}(20,15)
\vsixrangefourl
\nnint{vm1}{vem1}
\fmf{plain,tension=1,left=0,width=1mm}{v4,ved1}
\end{fmfchar*}}}
\to\frac{\lambda^3\hat\lambda}{16}
\Big(-\frac{1}{8\varepsilon^2}+\frac{1}{2\varepsilon}\Big)\pone{1}
\col\\
S_{\mathbf{v}1}
&=
\settoheight{\eqoff}{$\times$}%
\setlength{\eqoff}{0.5\eqoff}%
\addtolength{\eqoff}{-8.5\unitlength}%
\raisebox{\eqoff}{%
\fmfframe(0,1)(-5,1){%
\begin{fmfchar*}(20,15)
\fmftop{v3}
\fmfbottom{v4}
\fmfforce{(0.125w,h)}{v3}
\fmfforce{(0.125w,0)}{v4}
\fmffixed{(0.25w,0)}{v2,v1}
\fmffixed{(0.25w,0)}{v3,v2}
\fmffixed{(0.25w,0)}{v4,v5}
\fmffixed{(0.25w,0)}{v5,v6}
\vsix{v1}{v2}{v3}{v4}{v5}{v6}
\nnint{vo4}{vo6}
\fmf{plain,tension=1,left=0,width=1mm}{v4,v6}
\end{fmfchar*}}}
\to\frac{(\lambda\hat\lambda)^2}{16}\frac{1}{\varepsilon}\Big(2-\frac{\pi^2}{6}\Big)\pone{1}
\col\\
S_{\mathbf{v}2}
&=
\settoheight{\eqoff}{$\times$}%
\setlength{\eqoff}{0.5\eqoff}%
\addtolength{\eqoff}{-8.5\unitlength}%
\raisebox{\eqoff}{%
\fmfframe(0,1)(-5,1){%
\begin{fmfchar*}(20,15)
\fmftop{v3}
\fmfbottom{v4}
\fmfforce{(0.125w,h)}{v3}
\fmfforce{(0.125w,0)}{v4}
\fmffixed{(0.25w,0)}{v2,v1}
\fmffixed{(0.25w,0)}{v3,v2}
\fmffixed{(0.25w,0)}{v4,v5}
\fmffixed{(0.25w,0)}{v5,v6}
\vsix{v1}{v2}{v3}{v4}{v5}{v6}
\nnint{vm3}{vm2}
\fmf{plain,tension=1,left=0,width=1mm}{v4,v6}
\end{fmfchar*}}}
\to-\frac{\lambda^3\hat\lambda}{16}\frac{1}{4\varepsilon^2}\pone{1}
\col\\
S_{\mathbf{v}3}
&=
\settoheight{\eqoff}{$\times$}%
\setlength{\eqoff}{0.5\eqoff}%
\addtolength{\eqoff}{-8.5\unitlength}%
\raisebox{\eqoff}{%
\fmfframe(0,1)(-5,1){%
\begin{fmfchar*}(20,15)
\fmftop{v3}
\fmfbottom{v4}
\fmfforce{(0.125w,h)}{v3}
\fmfforce{(0.125w,0)}{v4}
\fmffixed{(0.25w,0)}{v2,v1}
\fmffixed{(0.25w,0)}{v3,v2}
\fmffixed{(0.25w,0)}{v4,v5}
\fmffixed{(0.25w,0)}{v5,v6}
\vsix{v1}{v2}{v3}{v4}{v5}{v6}
\nnint{vm3}{vm4}
\fmf{plain,tension=1,left=0,width=1mm}{v4,v6}
\end{fmfchar*}}}
\to\frac{\lambda\hat\lambda^3}{16}
\Big(-\frac{1}{8\varepsilon^2}-\frac{3\pi^2}{8\varepsilon}\Big)\pone{1}
\col\\
S_{\mathbf{v}4}
&=
\settoheight{\eqoff}{$\times$}%
\setlength{\eqoff}{0.5\eqoff}%
\addtolength{\eqoff}{-8.5\unitlength}%
\raisebox{\eqoff}{%
\fmfframe(0,1)(-5,1){%
\begin{fmfchar*}(20,15)
\fmftop{v3}
\fmfbottom{v4}
\fmfforce{(0.125w,h)}{v3}
\fmfforce{(0.125w,0)}{v4}
\fmffixed{(0.25w,0)}{v2,v1}
\fmffixed{(0.25w,0)}{v3,v2}
\fmffixed{(0.25w,0)}{v4,v5}
\fmffixed{(0.25w,0)}{v5,v6}
\vsix{v1}{v2}{v3}{v4}{v5}{v6}
\nnint{vo4}{vo5}
\fmf{plain,tension=1,left=0,width=1mm}{v4,v6}
\end{fmfchar*}}}
\to\frac{\lambda\hat\lambda}{16}
\Big[\lambda\hat\lambda
\frac{1}{\varepsilon}\Big(8-\frac{2}{3}\pi^2\Big)
+\lambda^2
\Big(-\frac{1}{4\varepsilon^2}
+\frac{1}{\varepsilon}\Big(1-\frac{7}{12}\pi^2\Big)\Big)
\Big]\pone{1}
\col\\
S_{\mathbf{s}1}
&=
\settoheight{\eqoff}{$\times$}%
\setlength{\eqoff}{0.5\eqoff}%
\addtolength{\eqoff}{-8.5\unitlength}%
\raisebox{\eqoff}{%
\fmfframe(0,1)(-5,1){%
\begin{fmfchar*}(20,15)
\fmftop{v3}
\fmfbottom{v4}
\fmfforce{(0.125w,h)}{v3}
\fmfforce{(0.125w,0)}{v4}
\fmffixed{(0.25w,0)}{v2,v1}
\fmffixed{(0.25w,0)}{v3,v2}
\fmffixed{(0.25w,0)}{v4,v5}
\fmffixed{(0.25w,0)}{v5,v6}
\vsix{v1}{v2}{v3}{v4}{v5}{v6}
\vacpol{v3}{vc6}
\fmf{plain,tension=1,left=0,width=1mm}{v4,v6}
\end{fmfchar*}}}
\to\frac{\lambda\hat\lambda}{16}\Big[
\lambda\hat\lambda\frac{3}{2\varepsilon^2}
+(\lambda-\hat\lambda)^2\frac{1}{4\varepsilon^2}\Big]\pone{1}
\col\\
S_{\mathbf{s}2}
&=
\settoheight{\eqoff}{$\times$}
\setlength{\eqoff}{0.5\eqoff}%
\addtolength{\eqoff}{-8.5\unitlength}%
\smash[b]{%
\raisebox{\eqoff}{%
\fmfframe(0,1)(-5,1){%
\begin{fmfchar*}(20,15)
\fmftop{v3}
\fmfbottom{v4}
\fmfforce{(0.125w,h)}{v3}
\fmfforce{(0.125w,0)}{v4}
\fmffixed{(0.25w,0)}{v2,v1}
\fmffixed{(0.25w,0)}{v3,v2}
\fmffixed{(0.25w,0)}{v4,v5}
\fmffixed{(0.25w,0)}{v5,v6}
\vsix{v1}{v2}{v3}{v4}{v5}{v6}
\vacpol{vc6}{v4}
\fmf{plain,tension=1,left=0,width=1mm}{v4,v6}
\end{fmfchar*}}}}
\to\frac{\lambda\hat\lambda}{16}
\Big[\lambda\hat\lambda\Big(\frac{3}{4\varepsilon^2}
-\frac{1}{\varepsilon}\Big(11-\frac{3}{4}\pi^2\Big)\Big)
+(\lambda-\hat\lambda)^2\Big(\frac{1}{8\varepsilon^2}
-\frac{1}{\varepsilon}\Big(\frac{1}{2}-\frac{1}{8}\pi^2\Big)\Big)
\Big]\pone{1}
\pnt
\end{aligned}
\end{equation}
The reflected diagrams for the first two diagrams are constructed according 
to the transformations in appendix \ref{app:symmetries}. The 
diagrams $S_{\mathbf{v}1}$ to $S_{\mathbf{v}4}$ then contribute with a
factor of two. The diagrams $S_{\mathbf{s}1}$ contributes with a factor
$\frac{3}{2}$, since there are three external fields which can obtain
a self-energy correction which counts with factor $\frac{1}{2}$ in the
renormalization of the composite operator. Finally, $S_{\mathbf{s}2}$
contributes with a factor of three. 
We find for the sum
\begin{equation}\label{Sn}
\begin{aligned}
S_{\mathbf{n}}
&=\frac{\lambda\hat\lambda}{16}
\Big[
\lambda\hat\lambda\Big(
\frac{11}{4\varepsilon^2}
-\frac{1}{\varepsilon}\Big(12+\frac{23}{12}\pi^2\Big)\Big)
+(\lambda-\hat\lambda)^2\Big(
-\frac{1}{8\varepsilon^2}-\frac{23}{24\varepsilon}\pi^2\Big)
\Big]\pone{1}
\pnt
\end{aligned}
\end{equation}

\subsection{Two-loop renormalization of the scalar six-vertex}
\label{subsec:V6renormalization}

The two-loop vertex renormalization of the scalar six-vertex allows 
us to fix a sign discrepancy in the literature\footnote{The antisymmetric parts in the product of two $\gamma$ matrices \eqref{gammaprod} in \cite{Benna:2008zy} and \cite{Bak:2008cp} differ by a sign.} parameterized 
by $z=\pm1$ which affects the four-loop diagrams with
fermion triangles, c.f.\ \eqref{Ftgraphs} and \eqref{Ft}.
Neglecting a factor $i^2=-1$ for the Wick rotation, 
the renormalization of the six-scalar vertex is found as
\begin{equation}
\begin{aligned}
\settoheight{\eqoff}{$\times$}%
\setlength{\eqoff}{0.5\eqoff}%
\addtolength{\eqoff}{-5\unitlength}%
\smash[b]{%
\raisebox{\eqoff}{%
\fmfframe(0,0)(0,0){%
\begin{fmfchar*}(10,10)
\fmftop{v1}
\fmfbottom{v4}
\fmfforce{(0.5w,h)}{v1}
\fmfforce{(0.5w,0)}{v4}
\fmfpoly{phantom}{v1,v2,v3,v4,v5,v6}
\fmf{plain}{v1,vc1}
\fmf{plain}{vc1,v2}
\fmf{plain}{v3,vc1}
\fmf{plain}{vc1,v4}
\fmf{plain}{v5,vc1}
\fmf{plain}{v6,vc1}
\fmffreeze
\fmfposition
\fmfv{decor.shape=circle,decor.filled=full,decor.size=0.25h}{vc1}
\end{fmfchar*}}}}
&=
\frac{6}{2}
\settoheight{\eqoff}{$\times$}%
\setlength{\eqoff}{0.5\eqoff}%
\addtolength{\eqoff}{-5\unitlength}%
\smash[b]{%
\raisebox{\eqoff}{%
\fmfframe(0,0)(0,0){%
\begin{fmfchar*}(10,10)
\fmftop{v1}
\fmfbottom{v4}
\fmfforce{(0.5w,h)}{v1}
\fmfforce{(0.5w,0)}{v4}
\fmfpoly{phantom}{v1,v2,v3,v4,v5,v6}
\fmf{plain}{v1,vc1}
\fmf{plain}{vc1,v2}
\fmf{plain}{v3,vc1}
\fmf{plain}{vc1,v4}
\fmf{plain}{v5,vc1}
\fmf{plain}{v6,vc1}
\fmffreeze
\fmfposition
\vacpol{v1}{vc1}
\end{fmfchar*}}}}
+3\Bigg(
\settoheight{\eqoff}{$\times$}%
\setlength{\eqoff}{0.5\eqoff}%
\addtolength{\eqoff}{-5\unitlength}%
\smash[b]{%
\raisebox{\eqoff}{%
\fmfframe(0,0)(0,0){%
\begin{fmfchar*}(10,10)
\fmftop{v1}
\fmfbottom{v4}
\fmfforce{(0.5w,h)}{v1}
\fmfforce{(0.5w,0)}{v4}
\fmfpoly{phantom}{v1,v2,v3,v4,v5,v6}
\fmf{plain}{v1,vc1}
\fmf{plain}{vc1,v2}
\fmf{plain}{v3,vc1}
\fmf{plain}{vc1,v4}
\fmf{plain}{v5,vc1}
\fmf{plain}{v6,vc1}
\fmffreeze
\fmf{phantom}{v1,ve1}
\fmf{phantom}{ve1,vc1}
\fmf{phantom}{v2,ve2}
\fmf{phantom}{ve2,vc1}
\fmfposition
\nnint{vloc(__ve1)}{vloc(__ve2)}
\end{fmfchar*}}}}
+
\settoheight{\eqoff}{$\times$}%
\setlength{\eqoff}{0.5\eqoff}%
\addtolength{\eqoff}{-5\unitlength}%
\smash[b]{%
\raisebox{\eqoff}{%
\fmfframe(0,0)(0,0){%
\begin{fmfchar*}(10,10)
\fmftop{v1}
\fmfbottom{v4}
\fmfforce{(0.5w,h)}{v1}
\fmfforce{(0.5w,0)}{v4}
\fmfpoly{phantom}{v1,v2,v3,v4,v5,v6}
\fmf{plain}{v1,vc1}
\fmf{plain}{vc1,v2}
\fmf{plain}{v3,vc1}
\fmf{plain}{vc1,v4}
\fmf{plain}{v5,vc1}
\fmf{plain}{v6,vc1}
\fmffreeze
\fmf{phantom}{v1,ve1}
\fmf{phantom}{ve1,vc1}
\fmf{phantom}{v6,ve2}
\fmf{phantom}{ve2,vc1}
\fmfposition
\nnint{vloc(__ve1)}{vloc(__ve2)}
\end{fmfchar*}}}}
\Bigg)
+3
\settoheight{\eqoff}{$\times$}%
\setlength{\eqoff}{0.5\eqoff}%
\addtolength{\eqoff}{-5\unitlength}%
\smash[b]{%
\raisebox{\eqoff}{%
\fmfframe(0,0)(0,0){%
\begin{fmfchar*}(10,10)
\fmftop{v1}
\fmfbottom{v4}
\fmfforce{(0.5w,h)}{v1}
\fmfforce{(0.5w,0)}{v4}
\fmfpoly{phantom}{v1,v2,v3,v4,v5,v6}
\fmffixed{(whatever,0)}{vc1,v2}
\fmffixed{(whatever,0)}{vc2,v3}
\fmf{plain}{v1,vc1}
\fmf{plain}{vc1,v2}
\fmf{plain}{v3,vc2}
\fmf{plain}{vc2,v4}
\fmf{plain,tension=0.333,left=0.5}{vc1,vc2}
\fmf{plain,tension=0.333}{vc1,vc2}
\fmf{plain,tension=0.333,right=0.5}{vc1,vc2}
\fmf{plain}{v5,vc2}
\fmf{plain}{v6,vc1}
\end{fmfchar*}}}}\\
&\phantom{{}={}}
+3\Bigg(
\settoheight{\eqoff}{$\times$}%
\setlength{\eqoff}{0.5\eqoff}%
\addtolength{\eqoff}{-5\unitlength}%
\smash[b]{%
\raisebox{\eqoff}{%
\fmfframe(0,0)(0,0){%
\begin{fmfchar*}(10,10)
\fmftop{v1}
\fmfbottom{v4}
\fmfforce{(0.5w,h)}{v1}
\fmfforce{(0.5w,0)}{v4}
\fmfpoly{phantom}{v1,v2,v3,v4,v5,v6}
\fmffixed{(whatever,0)}{vc1,v2}
\fmffixed{(whatever,0)}{vc2,v3}
\fmfpoly{phantom}{vc1,vc2,vc3}
\fmf{plain}{v1,vc1}
\fmf{plain}{vc1,v2}
\fmf{plain}{v3,vc2}
\fmf{plain}{vc2,v4}
\fmf{plain}{v5,vc3}
\fmf{plain}{vc3,v6}
\fmf{dashes}{vc1,vc2}
\fmf{dashes}{vc2,vc3}
\fmf{dashes}{vc3,vc1}
\fmffreeze
\fmfposition
\vacpol{vc1}{vc2}
\end{fmfchar*}}}}
+
\settoheight{\eqoff}{$\times$}%
\setlength{\eqoff}{0.5\eqoff}%
\addtolength{\eqoff}{-5\unitlength}%
\smash[b]{%
\raisebox{\eqoff}{%
\fmfframe(0,0)(0,0){%
\begin{fmfchar*}(10,10)
\fmftop{v1}
\fmfbottom{v4}
\fmfforce{(0.5w,h)}{v1}
\fmfforce{(0.5w,0)}{v4}
\fmfpoly{phantom}{v1,v2,v3,v4,v5,v6}
\fmffixed{(whatever,0)}{vc2,v5}
\fmffixed{(whatever,0)}{vc3,v6}
\fmfpoly{phantom}{vc1,vc2,vc3}
\fmf{plain}{v2,vc1}
\fmf{plain}{vc1,v3}
\fmf{plain}{v4,vc2}
\fmf{plain}{vc2,v5}
\fmf{plain}{v6,vc3}
\fmf{plain}{vc3,v1}
\fmf{dashes}{vc1,vc2}
\fmf{dashes}{vc2,vc3}
\fmf{dashes}{vc3,vc1}
\fmffreeze
\fmfposition
\vacpol{vc1}{vc2}
\end{fmfchar*}}}}
\Bigg)
+3\Bigg(
\settoheight{\eqoff}{$\times$}%
\setlength{\eqoff}{0.5\eqoff}%
\addtolength{\eqoff}{-5\unitlength}%
\smash[b]{%
\raisebox{\eqoff}{%
\fmfframe(0,0)(0,0){%
\begin{fmfchar*}(10,10)
\fmftop{v1}
\fmfbottom{v4}
\fmfforce{(0.5w,h)}{v1}
\fmfforce{(0.5w,0)}{v4}
\fmfpoly{phantom}{v1,v2,v3,v4,v5,v6}
\fmffixed{(whatever,0)}{vc1,v2}
\fmffixed{(whatever,0)}{vc2,v3}
\fmfpoly{phantom}{vc1,vc2,vc3}
\fmf{plain}{v1,vc1}
\fmf{plain}{vc1,v2}
\fmf{plain}{v3,vc2}
\fmf{plain}{vc2,v4}
\fmf{plain}{v5,vc3}
\fmf{plain}{vc3,v6}
\fmf{dashes}{vc1,vtm1}
\fmf{dashes}{vtm1,vc2}
\fmf{dashes}{vc2,vtm2}
\fmf{dashes}{vtm2,vc3}
\fmf{dashes}{vc3,vtm3}
\fmf{dashes}{vtm3,vc1}
\fmffreeze
\fmfposition
\fmf{photon}{vtm1,vtm2}
\end{fmfchar*}}}}
+
\settoheight{\eqoff}{$\times$}%
\setlength{\eqoff}{0.5\eqoff}%
\addtolength{\eqoff}{-5\unitlength}%
\smash[b]{%
\raisebox{\eqoff}{%
\fmfframe(0,0)(0,0){%
\begin{fmfchar*}(10,10)
\fmftop{v1}
\fmfbottom{v4}
\fmfforce{(0.5w,h)}{v1}
\fmfforce{(0.5w,0)}{v4}
\fmfpoly{phantom}{v1,v2,v3,v4,v5,v6}
\fmffixed{(whatever,0)}{vc2,v5}
\fmffixed{(whatever,0)}{vc3,v6}
\fmfpoly{phantom}{vc1,vc2,vc3}
\fmf{plain}{v2,vc1}
\fmf{plain}{vc1,v3}
\fmf{plain}{v4,vc2}
\fmf{plain}{vc2,v5}
\fmf{plain}{v6,vc3}
\fmf{plain}{vc3,v1}
\fmf{dashes}{vc1,vtm1}
\fmf{dashes}{vtm1,vc2}
\fmf{dashes}{vc2,vtm2}
\fmf{dashes}{vtm2,vc3}
\fmf{dashes}{vc3,vtm3}
\fmf{dashes}{vtm3,vc1}
\fmffreeze
\fmfposition
\fmf{photon}{vtm1,vtm2}
\end{fmfchar*}}}}
\Bigg)
+6
\settoheight{\eqoff}{$\times$}%
\setlength{\eqoff}{0.5\eqoff}%
\addtolength{\eqoff}{-5\unitlength}%
\smash[b]{%
\raisebox{\eqoff}{%
\fmfframe(0,0)(0,0){%
\begin{fmfchar*}(10,10)
\fmftop{v1}
\fmfbottom{v4}
\fmfforce{(0.5w,h)}{v1}
\fmfforce{(0.5w,0)}{v4}
\fmfpoly{phantom}{v1,v2,v3,v4,v5,v6}
\fmffixed{(whatever,0)}{vc1,v2}
\fmffixed{(whatever,0)}{vc2,v3}
\fmf{plain}{v1,vc1}
\fmf{plain}{vc1,v2}
\fmf{plain}{v3,vc2}
\fmf{plain}{vc2,v4}
\fmf{plain}{v5,vca3}
\fmf{plain}{vcb3,v6}
\fmf{dashes}{vc1,vc2}
\fmf{dashes}{vc2,vca3}
\fmf{plain,left=0.5}{vca3,vcb3}
\fmf{dashes}{vcb3,vc1}
\fmffreeze
\fmfposition
\fmf{dashes,right=0.5}{vca3,vcb3}
\end{fmfchar*}}}}
+6
\settoheight{\eqoff}{$\times$}%
\setlength{\eqoff}{0.5\eqoff}%
\addtolength{\eqoff}{-5\unitlength}%
\smash[b]{%
\raisebox{\eqoff}{%
\fmfframe(0,0)(0,0){%
\begin{fmfchar*}(10,10)
\fmftop{v1}
\fmfbottom{v4}
\fmfforce{(0.5w,h)}{v1}
\fmfforce{(0.5w,0)}{v4}
\fmfpoly{phantom}{v1,v2,v3,v4,v5,v6}
\fmffixed{(whatever,0)}{vc1,v2}
\fmffixed{(whatever,0)}{vc2,v3}
\fmf{plain}{v1,vc1}
\fmf{plain}{vc1,v2}
\fmf{plain}{v3,vc2}
\fmf{plain}{vc2,v4}
\fmf{plain}{v5,vca3}
\fmf{plain}{vcb3,v6}
\fmf{dashes}{vc1,vc2}
\fmf{dashes}{vc2,vca3}
\fmf{dashes,left=0.5}{vca3,vcb3}
\fmf{dashes}{vcb3,vc1}
\fmffreeze
\fmfposition
\fmf{plain,right=0.5}{vca3,vcb3}
\end{fmfchar*}}}}
+3
\settoheight{\eqoff}{$\times$}%
\setlength{\eqoff}{0.5\eqoff}%
\addtolength{\eqoff}{-5\unitlength}%
\smash[b]{%
\raisebox{\eqoff}{%
\fmfframe(0,0)(0,0){%
\begin{fmfchar*}(10,10)
\fmftop{v1}
\fmfbottom{v4}
\fmfforce{(0.5w,h)}{v1}
\fmfforce{(0.5w,0)}{v4}
\fmfpoly{phantom}{v1,v2,v3,v4,v5,v6}
\fmf{plain}{v1,vc1}
\fmf{plain}{vc2,v2}
\fmf{plain}{vc3,v4}
\fmf{plain}{v3,vc3}
\fmf{plain}{v5,vc4}
\fmf{plain}{vc1,v6}
\fmf{dashes}{vc1,vc2}
\fmf{dashes}{vc2,vc3}
\fmf{dashes}{vc3,vc4}
\fmf{dashes}{vc4,vc1}
\fmffreeze
\fmfposition
\fmf{plain}{vc2,vc4}
\end{fmfchar*}}}}\\
&=\lambda\hat\lambda\frac{3}{2\varepsilon}i(z-1)\pone{1}
\pnt
\end{aligned}
\end{equation} 
Non-renormalization of the coupling and hence superconformal invariance requires $z=1$, which corresponds to the sign choice in \cite{Benna:2008zy}.

\section{Result}

The four-loop term of the
renormalization constant of the composite operators
\begin{equation}
\mathcal{O}_{a,\text{ren}}=\mathcal{Z}_{a}{}^b\mathcal{O}_{b,\text{bare}}
\col\qquad
\mathcal{Z}=\unitmatrix+\bar\lambda^2\mathcal{Z}_2+\bar\lambda^4\mathcal{Z}_4+\dots
\label{opren}
\end{equation}
is given as the negative of the sum of $S$,
$F_{\mathbf{s}}$, $F_{\mathbf{t}}$, $S_{\mathbf{n}}$ in equations 
\eqref{S}, \eqref{Fs}, \eqref{Ft}, \eqref{Sn}.
Fixing $z=1$ as determined in subsection \ref{subsec:V6renormalization},
we obtain
\begin{equation}\label{Z4}
\begin{aligned}
\mathcal{Z}_4
&\to\frac{1}{16}
\Big[\Big(\frac{1}{2\varepsilon^2}-\frac{2}{\varepsilon}\Big)(\ptwo{1}{3}+\ptwo{3}{1})
+\frac{1}{\varepsilon^2}\ptwo{1}{2}\\
&\phantom{{}={}\frac{1}{16}\Big[}
+\Big(-\frac{15}{4\varepsilon^2}+\frac{1}{\varepsilon}\Big(12+\frac{4}{3}\pi^2\Big)
+\sigma^2\Big(\frac{1}{8\varepsilon^2}+\frac{\pi^2}{3\varepsilon}\Big)
\Big)\pone{1}\Big]\pnt
\end{aligned}
\end{equation}
The dilatation operator is then obtained from the renormalization
constant $\mathcal{Z}$ as
\begin{equation}\label{anomdim}
D=\lim_{\varepsilon\rightarrow0}\left[2\varepsilon\bar\lambda
\frac{\de}{\de\bar\lambda}\ln\mathcal{Z}(\bar\lambda,\varepsilon)\right]\pnt
\end{equation}
Effectively, the above definition extracts the coefficient of the 
$\frac{1}{\varepsilon}$ pole at a given loop order $L$ and multiplies it 
by a factor $2L$. 
The four-loop dilatation operator for odd sites 
is thus the coefficient of the $\frac{1}{\varepsilon}$
pole of \eqref{Z4} multiplied by $8$. We  must still add the 
neglected identity part, which we fix by demanding that the ground state has
zero eigenvalue. We thus obtain 
\begin{equation}\label{D4odd}
\begin{aligned}
D_{4,\text{odd}}
&=-(4+4\zeta(2)
+\sigma^2\zeta(2))\pone{}
+(6+4\zeta(2)
+\sigma^2\zeta(2))\pone{1}-\ptwo{1}{3}-\ptwo{3}{1}\col
\end{aligned}
\end{equation}
where $\zeta(2)=\frac{\pi^2}{6}$.
By comparing this result with \eqref{D4}, we immediately find 
that the four-loop term of the function
$h^2(\lambda)$ in \eqref{h4ansatz} is given by
\begin{equation}
h_4(\sigma)=-4\zeta(2)-\sigma^2\zeta(2)
\pnt
\end{equation}

\subsection{An interesting limit}

The presence of the extra parameter $\sigma$ suggests an interesting limit.
Suppose that $\lambda\gg\hat\lambda$ while $\lambda$ is still in the 
perturbative regime. This corresponds to a dominance of the term
$\sigma^2h_{4,\sigma}$ w.r.t.\ $h_4$ at four loops.
Letting $\mathcal{D}=\hat\lambda^{-1}(D-2L)$, then $\mathcal{D}$ in this
limit reduces to 
\begin{equation}\label{localham}
\begin{aligned}
\mathcal{D}=\lambda(1-\zeta(2)\lambda^2)
(2\pone{}-\pone{1}-\pone{2})
\end{aligned}
\end{equation}
for the  $SU(2)\times SU(2)$ sector.
In other words, $\mathcal{D}$ is the Hamiltonian for two Heisenberg spin
chains! In fact, at higher loops, it is clear that the dominant
contribution also comes from the next-to-nearest neighbour interactions 
with the same permutation structure, hence
in this limit $\mathcal{D}$ is the same local Hamiltonian as in
(\ref{localham}), except with the $\la$ dependent prefactor replaced
with an all-loop function $f(\la)$. 

Whether or not the integrability
holds outside of the $SU(2)\times SU(2)$ sector to four
loops or higher remains to be seen.  It is also true that this scaling only
applies to the anomalous dimension and not to the full dimension of
the operator.  Added to this, there are reasons to believe that the
ABJ Chern-Simons theory is  inconsistent if $\lambda-\hat\lambda>1$
\cite{Aharony:2008gk}, 
thus this limit is likely to fail at strong coupling.


\section{Four-loop wrapping interactions}

To find the correct anomalous dimensions of a length four state
at four loops, we have to consider the wrapping interactions
\cite{Beisert:2004hm,Sieg:2005kd}. 
Recently, Gromov, Kazakov and Vieira have made a prediction for the
four-loop wrapping contribution for  scalar operators in the {\bf 20}
representation of $SU(4)$ \cite{Gromov:2009tv}.  
The highest weight of
this representation is in the $SU(2)\times SU(2)$ sector. 
These
authors find their result by  applying the thermodynamic Bethe ansatz
(TBA), first introduced for the original $\AdS/\text{CFT}$ 
correspondence in \cite{Ambjorn:2005wa,Arutyunov:2007tc}, 
using the predicted asymptotic Bethe equations
\cite{Gromov:2008qe}.  In particular, based on the string hypothesis for the mirror theory \cite {Arutyunov:2009zu}, the TBA is formulated in terms
of a $Y$-system \cite{Gromov:2009bc,Bombardelli:2009ns,Arutyunov:2009ur,Bombardelli:2009xz,Gromov:2009at}, 
a series of difference equations, that 
lend themselves to an efficient order by order solution.

Following the strategy of 
\cite{Fiamberti:2008sh}, we first subtract from the asymptotic
dilatation operator \eqref{D4} the range five interactions (given by
$S_2$ and the reflected diagram). This
leaves for $D_{4,\text{odd}}$ in \eqref{D4odd}
\begin{equation}\label{D4oddsub}
\begin{aligned}
D_{4,\text{odd}}^\text{sub}
&=(4+4\zeta(2)
+\sigma^2\zeta(2))(\pone{1}-\pone{})\pnt
\end{aligned}
\end{equation}
We then have to add the following wrapping diagrams which contain a permutation structure
\begin{equation}
\begin{aligned}
W_1&=
\settoheight{\eqoff}{$\times$}%
\setlength{\eqoff}{0.5\eqoff}%
\addtolength{\eqoff}{-10\unitlength}%
\raisebox{\eqoff}{%
\fmfframe(2,1)(2,4){%
\begin{fmfchar*}(20,15)
\fmftop{v5}
\fmfbottom{v6}
\fmfforce{(-0.125w,h)}{v5}
\fmfforce{(-0.125w,0)}{v6}
\fmffixed{(0.25w,0)}{v2,v1}
\fmffixed{(0.25w,0)}{v3,v2}
\fmffixed{(0.25w,0)}{v4,v3}
\fmffixed{(0.25w,0)}{v5,v4}
\fmffixed{(0.25w,0)}{v6,v7}
\fmffixed{(0.25w,0)}{v7,v8}
\fmffixed{(0.25w,0)}{v8,v9}
\fmffixed{(0.25w,0)}{v9,v10}
\fmffixed{(-0.125w,0)}{v10,vb}
\fmffixed{(-0.125w,0)}{va,v7}
\fmffixed{(whatever,0)}{v61,v62}
\fmf{phantom,left=0.25}{v1,v62}
\fmf{plain}{v2,v62}
\fmf{phantom,right=0.25}{v3,v62}
\fmf{plain,left=0.25}{v8,v62}
\fmf{plain}{v9,v62}
\fmf{plain,right=0.25}{v10,v62}
\fmf{phantom,left=0.25}{v6,v61}
\fmf{plain}{v7,v61}
\fmf{phantom,right=0.25}{v8,v61}
\fmf{plain,left=0.25}{v3,v61}
\fmf{plain}{v4,v61}
\fmf{plain,right=0.25}{v5,v61}
\fmffreeze
\fmf{plain}{v61,v62}
\fmf{plain,tension=1,left=0,width=1mm}{v7,v10}
\plainwrap{v61}{v7}{v10}{v62}
\end{fmfchar*}}}
\to 
\frac{(\lambda\hat\lambda)^2}{16}\Big(-\frac{1}{2\varepsilon^2}+\frac{2}{\varepsilon}\Big)
(\ptwo{1}{3}-\pone{1})
\col\\
W_2&=
\settoheight{\eqoff}{$\times$}%
\setlength{\eqoff}{0.5\eqoff}%
\addtolength{\eqoff}{-10\unitlength}%
\raisebox{\eqoff}{%
\fmfframe(2,1)(2,4){%
\begin{fmfchar*}(20,15)
\fmftop{v4}
\fmfbottom{v5}
\fmfforce{(0.125w,h)}{v4}
\fmfforce{(0.125w,0)}{v5}
\fmffixed{(0.25w,0)}{v2,v1}
\fmffixed{(0.25w,0)}{v3,v2}
\fmffixed{(0.25w,0)}{v4,v3}
\fmffixed{(0.25w,0)}{v5,v6}
\fmffixed{(0.25w,0)}{v6,v7}
\fmffixed{(0.25w,0)}{v7,v8}
\fmffixed{(whatever,0)}{v61,v62}
\fmf{plain,left=0.25}{v1,v62}
\fmf{plain,right=0.25}{v2,v62}
\fmf{plain,left=0.25}{v7,v62}
\fmf{plain,right=0.25}{v8,v62}
\fmf{plain,left=0.25}{v5,v61}
\fmf{plain,right=0.25}{v6,v61}
\fmf{plain,left=0.25}{v3,v61}
\fmf{plain,right=0.25}{v4,v61}
\fmffreeze
\fmf{plain,left=0}{v61,v62}
\fmf{plain,tension=1,left=0,width=1mm}{v5,v8}
\plainwrap{v61}{v5}{v8}{v62}
\end{fmfchar*}}}
\to
\frac{(\lambda\hat\lambda)^2}{16}\frac{\pi^2}{2\varepsilon}\pone{1}
\col\\
%
%
W_3&=
\settoheight{\eqoff}{$\times$}%
\setlength{\eqoff}{0.5\eqoff}%
\addtolength{\eqoff}{-10\unitlength}%
\raisebox{\eqoff}{%
\fmfframe(2,1)(3,4){%
\begin{fmfchar*}(20,15)
\fmftop{v5}
\fmfbottom{v6}
\fmfforce{(0.125w,h)}{v5}
\fmfforce{(0.125w,0)}{v6}
\fmffixed{(0.25w,0)}{v2,v1}
\fmffixed{(0.25w,0)}{v3,v2}
\fmffixed{(0.25w,0)}{v4,v3}
\fmffixed{(0.25w,0)}{v5,v4}
\fmffixed{(0.25w,0)}{v6,v7}
\fmffixed{(0.25w,0)}{v7,v8}
\fmffixed{(0.25w,0)}{v8,v9}
\fmffixed{(0.25w,0)}{v9,v10}
\fmffixed{(whatever,0)}{v61,v62}
\fmffixed{(whatever,0)}{v61,v6a}
\fmffixed{(whatever,0)}{v62,v6b}
\fmf{plain,left=0.25}{v1,v6b}
\fmf{plain,left=0.25}{v2,v62}
\fmf{plain,left=0.25}{v8,v62}
\fmf{plain,left=0.25}{v9,v6b}
\fmf{plain,left=0.25}{v6,v61}
\fmf{plain,right=0.25}{v7,v61}
\fmf{plain,left=0.25}{v3,v6a}
\fmf{plain,right=0.25}{v4,v6a}
\fmffreeze
\fmf{plain,tension=1,left=0,width=1mm}{v6,v9}
\dasheswrap{v61}{v5}{v9}{v6b}
\fmfi{dashes}{vloc(__v61) ..vloc(__v6b)}
\end{fmfchar*}}}
\to
-\frac{(\lambda\hat\lambda)^2}{16}\frac{\pi^2}{4\varepsilon}2\pone{1}
\col\\
W_4&=
\settoheight{\eqoff}{$\times$}%
\setlength{\eqoff}{0.5\eqoff}%
\addtolength{\eqoff}{-10\unitlength}%
\raisebox{\eqoff}{%
\fmfframe(2,1)(3,4){%
\begin{fmfchar*}(20,15)
\fmftop{v5}
\fmfbottom{v6}
\fmfforce{(0.125w,h)}{v5}
\fmfforce{(0.125w,0)}{v6}
\fmffixed{(0.25w,0)}{v2,v1}
\fmffixed{(0.25w,0)}{v3,v2}
\fmffixed{(0.25w,0)}{v4,v3}
\fmffixed{(0.25w,0)}{v5,v4}
\fmffixed{(0.25w,0)}{v6,v7}
\fmffixed{(0.25w,0)}{v7,v8}
\fmffixed{(0.25w,0)}{v8,v9}
\fmffixed{(0.25w,0)}{v9,v10}
\fmffixed{(whatever,0)}{v61,v62}
\fmffixed{(whatever,0)}{v61,v6a}
\fmffixed{(whatever,0)}{v62,v6b}
\fmf{plain,left=0.25}{v1,v6b}
\fmf{plain,left=0.25}{v2,v62}
\fmf{plain,right=0.25}{v8,v6a}
\fmf{plain,left=0.25}{v9,v6b}
\fmf{plain,left=0.25}{v6,v61}
\fmf{plain,right=0.25}{v7,v61}
\fmf{plain,right=0.25}{v3,v62}
\fmf{plain,right=0.25}{v4,v6a}
\fmffreeze
\fmf{plain,tension=1,left=0,width=1mm}{v6,v9}
\dasheswrap{v61}{v5}{v9}{v6b}
\fmfi{dashes}{vloc(__v61) ..vloc(__v6b)}
\end{fmfchar*}}}
\to
\frac{(\lambda\hat\lambda)^2}{16}\frac{1}{\varepsilon}8\pone{1}
\col\\
W_5&=
\settoheight{\eqoff}{$\times$}%
\setlength{\eqoff}{0.5\eqoff}%
\addtolength{\eqoff}{-10\unitlength}%
\raisebox{\eqoff}{%
\fmfframe(2,1)(2,4){%
\begin{fmfchar*}(20,15)
\vsixrangefourl
\fmfi{photon}{vm6--vem6}
\fmf{plain,tension=1,left=0,width=1mm}{v4,ved1}
\wigglywrap{vm4}{v4}{ved1}{vem6}
\end{fmfchar*}}}
\to
\frac{(\lambda\hat\lambda)^2}{16}\frac{2}{\varepsilon}\Big(1-\frac{\pi^2}{12}\Big)\pone{1}
\pnt\\
\end{aligned}
\end{equation}
We then sum  the above diagrams,  where we include a factor of two for 
$W_3$ since its reflection  cannot be
mapped to itself by a cyclic rotation. To this sum we add the identity
terms needed to give zero wrapping for a chiral primary, remembering
that $\{1,3\}$ is equivalent to $\{\,\}$ for length four operators.
Taking the negative of the sum, 
extracting the residues of $\frac{1}{\epsilon}$ and multiplying by
 $8$ as in the previous section,  we find that the wrapping
contribution to the odd site dilatation operator 
is given by
\begin{equation}\label{D4oddw}
D_{4,\text{odd}}^\text{w}
=-(4-2\zeta(2))(\pone{1}-\pone{})
\pnt
\end{equation}
The {\bf 20} in the $SU(2)\times SU(2)$ sector is a completely antisymmetric state, so the eigenvalue of $\pone{1}-\pone{}$ is
$-8$.  Hence the four-loop wrapping contribution to the anomalous dimension is 
\begin{equation}
\gamma_{20}^\text{w}=(32-16\zeta(2))\bl^4\col
\end{equation}
which agrees  with the GKV $Y$-system prediction
\cite{Gromov:2009tv}. It does not depend on $\sigma^2$, which is 
expected from the topology of the diagrams and the fact that the
two-loop result does not depend on $\sigma^2$.
Adding $D_{4,\text{odd}}^\text{sub}$ to $D_{4,\text{odd}}^\text{w}$, 
we find 
that the anomalous dimension of the {\bf 20} is given by
\begin{equation}
\gamma_{\mathbf{20}}
=4+8\bar\lambda^2
-(48\zeta(2)+8\sigma^2\zeta(2))\bar\lambda^4
\col
\end{equation} 
where we have included the lower orders \cite{Minahan:2008hf}.
It is striking that the rational parts coming
from wrapping terms cancel the rational contribution in \eqref{D4oddsub}, 
leaving a maximally transcendental four-loop result for $\gamma_{20}$.


\section{Discussion}

In this paper we computed the four-loop correction to $h^2(\bl,\s)$.  The coefficients $h_4$ and $h_{4,\s}$ both have maximal transcendentality.  The significance of this is hard to say. It would be interesting to see if this pattern continues to higher loops.

In section 2 we argued  that the perturbative
expansion is invariant under parity to four-loop
  order in the coupling. Assuming  that our arguments can be extended to higher loops,  
$h^2(\bl,\s)$  would then have a perturbative expansion in even powers of $\s$. On the string world-sheet a nonzero $\s$ comes from turning on a  $b$ field flux through the 2-cycle of $\text{CP}^3$ \cite{Aharony:2008gk}.  This leads to nonperturbative corrections that do not directly affect  the strong coupling behavior of $h^2(\bl,\s)$.  
However, there is a change in  the radius of $\AdS_4$ and $\text{CP}^3$, which in the scaling limit is  \cite{Bergman:2009zh,Aharony:2008gk}\footnote{There could be a further correction to $R_s$ because of a
  half-integer shift of the $b$ field flux through the 2-cycle of
$\text{CP}^3$ \cite{Aharony:2009fc}. 
  We thank O. Bergman for
  discussions on this.}
\begin{equation}\label{Rs}
R_s^2=2^{5/2}\pi\sqrt{\mbox{min}(\la,\hla)-\frac{1}{24}+\frac12(\la-\hla)^2}\,,
\end{equation}
where $\mbox{min}(\la,\hla)$ is the minimum of $\la$ and $\hla$.
Hence, for $\bl\gg1$ $h^2(\bl,\s)$ has a cut in the $\s^2$ plane
with a branch point at $\s^2=0$.  To do a parity transformation, we
could smoothly deform $\s$ to $-\s$.    If we  circle the branch point
then $h^2$ is not invariant and parity would seem to be broken. From
the string theory analysis in \cite{Aharony:2008gk} it seems that the
correct thing to do is not to circle the branch
point, keeping parity
invariant.  It would be helpful to better understand what the physical
implications are of going to the other sheet.

We have also proposed a new local spin chain that arises when $\la\gg\hla$.  If this limit is integrable, then the Hamiltonian is guaranteed by the $\NN=6$ supersymmetry to have  the form in \cite{Zwiebel:2009vb,Minahan:2009te}, only differing by an overall function of $\la$.  It would be very interesting to check if it can be extended outside the $SU(2)\times SU(2)$ sector or to higher loops.  A feasible test would be in  the  scalar $SU(4)$ sector where integrability requires certain relative coefficients between nontrivial terms in the Hamiltonian.

It has been argued that the $U(M)_k\times U(N)_{-k}$ theory is nonunitary if $M>N+k$ \cite{Aharony:2008gk}, which corresponds to $\la-\hla>1$.  A possible signal of this in our limiting local spin chain would be the Hamiltonian changing sign at $\la=1$.  If this were to happen, then the non BPS operators would have negative anomalous dimensions, a clear violation of unitarity.  Since the sign of $h_{4,\s}$ is negative, this seems reasonably plausible.

One can also speculate on why $h^2(\bl,\s)$ is more complicated for
ABJ(M) than for $\NN=4$ SYM.  A possible clue is that $U(M)_k\times U(N)_{-k}$ is believed to be equivalent to $U(N)_k\times U(2N-M+k)_{-k}$ for $M>N$ \cite{Aharony:2008gk}. In terms of $\la$ and $\hla$ this corresponds to equivalence under $\la\to\hla$, $\hla\to 2\hla-\la+1$, which translates into rather complicated transformations for $\bl$ and $\s$ . If the ABJ theory is integrable, then under this transformation $h^2(\bl,\s)$ must be invariant.    From this alone we see that $h^2$ has to be a rather complicated function of both $\bl$ and $\s$.  Perhaps the asymptotic information for $h^2(\bl,\s)$ and the  invariance under these transformations will be enough to ultimately find the complete function.

We have also verified the GKV prediction for the wrapping
correction to the {\bf 20}.  This serves as an important check for the $Y$-system conjecture and also provides a useful consistency check for us. 
Finally, we note that as $h_4(\lambda)$ also the first wrapping correction is 
maximally transcendental.

\subsection*{Acknowledgments}
We would like to thank O.\ Bergman, T.\ Klose and K.\
Zarembo for very helpful discussions. 
The research of J.\ A.\ M.\ is supported in part by the
Swedish research council.  The research of C.\ S.\ is supported in part by
the European Marie Curie Research and Training Network 
ENRAGE (MRTN-CT-2004-005616).  J.\ A.\ M.\ thanks the
CTP at MIT and  the Galileo Institute in Florence  for kind
hospitality  during the course of this work. Support for these visits
comes from the STINT foundation and INFN. O.\ O. S.\  thanks NBI and  
C.\ S.\ thanks Uppsala University for hospitality during the course 
of this work.


\appendix

\section{Conventions}
\label{app:conventions}

In three dimensional spacetime with metric $\eta_{\mu\nu}=\diag(-,+,+)$
The antisymmetric tensor is normalized as $\epsilon^{012}=1$.
The product of $\gamma$-matrices is given by
\begin{equation}\label{gammaprod}
\begin{aligned}
\gamma_\mu\gamma_\nu
&=\eta_{\mu\nu}+z\epsilon_{\mu\nu\rho}\gamma^\rho
\col\\
\end{aligned}
\end{equation}
where $z=\pm1$ represents a sign.
The Clifford algebra is given by
\begin{equation}
\acomm{\gamma^\mu}{\gamma^\nu}=2\eta^{\mu\nu}
\begin{pmatrix} 1 & 0 \\ 0 & 1 \end{pmatrix}
\pnt
\end{equation}
The action reads
\begin{equation}
\begin{aligned}
S_\text{kin}&=\frac{k}{4\pi}\int\de^3x\tr\Big[
A_\alpha\Big(\epsilon^{\alpha\beta\gamma}\partial_\beta-\frac{1}{\zeta}\partial^\alpha\partial^\gamma\Big)A_\gamma
-\hat A_\alpha\Big(\epsilon^{\alpha\beta\gamma}\partial_\beta-\frac{1}{\zeta}\partial^\alpha\partial^\gamma\Big)\hat A_\gamma\\
&\phantom{{}={}\frac{k}{4\pi}\int\de^3x\tr\Big[}
+Y_A^\dagger\partial_\mu\partial^\mu Y^A+i\psi^{\dagger B}\dslash{\partial}\psi_B
+c^\ast\partial_\mu\partial^\mu c+\hat c^\ast\partial_\mu\partial^\mu \hat c
\Big]\\
S_\text{int}&=\frac{k}{4\pi}\int\de^3x\tr\Big[
\frac{2}{3}i\epsilon^{\alpha\beta\gamma}(A_\alpha A_\beta A_\gamma
-\hat A_\alpha\hat A_\beta\hat A_\gamma)\\
&\phantom{{}={}\frac{k}{4\pi}\int\de^3x\tr\Big[}
-iA_\mu Y^A\overset{\leftrightarrow}{\partial^\mu}Y_A^\dagger
-i\hat A_\mu Y_A^\dagger\overset{\leftrightarrow}{\partial^\mu}Y^A
+2Y_A^\dagger A_\mu Y^A\hat A^\mu\\
&\phantom{{}={}\frac{k}{4\pi}\int\de^3x\tr\Big[}
-\hat A_\mu\hat A_\mu Y_A^\dagger Y^A
-A_\mu A_\mu Y^AY_A^\dagger\\
&\phantom{{}={}\frac{k}{4\pi}\int\de^3x\tr\Big[}
-\psi^{\dagger B}\dslash A\psi_B
+\hat A_\mu\psi^{\dagger B}\gamma^\mu\psi_B
-iA^\mu\comm{c}{\partial_\mu c^\ast}
-i\hat A^\mu\comm{\hat c}{\partial_\mu\hat c^\ast}
\\
&\phantom{{}={}\frac{k}{4\pi}\int\de^3x\tr\Big[}
+\frac{1}{12}Y^AY_B^\dagger Y^CY_D^\dagger Y^EY_F^\dagger
(\delta_A^B\delta_C^D\delta_E^F
+\delta_A^F\delta_C^B\delta_E^D
-6\delta_A^B\delta_C^F\delta_E^D
+4\delta_A^D\delta_C^F\delta_E^B)\\
&\phantom{{}={}\frac{k}{4\pi}\int\de^3x\tr\Big[}
-\frac{i}{2}(
Y_A^\dagger Y^B\psi^{\dagger C}\psi_D
-\psi_D\psi^{\dagger C}Y^BY_A^\dagger)
(\delta_B^A\delta_C^D-2\delta_C^A\delta_B^D)\\
&\phantom{{}={}\frac{k}{4\pi}\int\de^3x\tr\Big[}
+\frac{i}{2}\epsilon^{ABCD}Y_A^\dagger\psi_BY_C^\dagger\psi_D
-\frac{i}{2}\epsilon_{ABCD}Y^A\psi^{\dagger B}Y^C\psi^{\dagger D}
\Big]
\pnt
\end{aligned}
\end{equation}
The Feynman rules are extracted from this action.

\section{Permutation structures}
\label{app:permstruc}

Since the operators in the $SU(2)\times SU(2)$ subsector are free of 
subtraces, the corresponding dilatation operator is given by an expansion
in terms of the permutation structures defined in \eqref{permstruc}.

The definition is similar as in the $\mathcal{N}=4$ SYM case,
\cite{Beisert:2003tq}, but each permutation in the product permutes neighbouring
fields at either odd or even sites. Also,
the permutation structures are shifted by two sides then inserted
along the chain and not by one site as in the $\mathcal{N}=4$ SYM
case.
Permutations at 
even and odd insertion points commute with each other, and only do not
commute if they have one position in common, i.e.\ if the integers in
the argument lists of the permutation structures differ by two.
The rules for manipulation which have to be modified compared to the
ones in the  $\mathcal{N}=4$ SYM case are
\begin{equation}
\begin{aligned}
\pfour{\dots}{a}{b}{\dots}&=\pfour{\dots}{b}{a}{\dots}\col\qquad
|a-b|\neq 2\col \\
\pthree{a}{\dots}{b}&=\pthree{a+2n}{\dots}{b+2n}\pnt
\end{aligned}
\end{equation}
We can shift all odd integers to the left and all even integers
in the argument lists to the right. The basis of permutation structures at
four loops hence reads
\begin{equation}
\pone{}\col\quad
\pone{1}\col\quad
\pone{2}\col\quad
\ptwo{1}{2}\col\quad
\ptwo{3}{2}\col\quad
\ptwo{1}{3}\col\quad
\ptwo{3}{1}\col\quad
\ptwo{2}{4}\col\quad
\ptwo{4}{2}\pnt
\end{equation}
The dilatation operator decomposes into three parts, acting on even, on 
odd and on mixed sides, respectively. These parts are defined by containing
permutation structures with respectively only odd, only even and both odd and even arguments.

\section{Symmetries}
\label{app:symmetries}

Let $c(\lambda,\hat\lambda)\pthree{a_1}{\dots}{a_m}$ 
be the result associated with a certain Feynman graph. 
We can easily construct the contribution from
the analogous graph shifted by one step along the chain of elementary fields 
from the corresponding operator. This exchanges all fields as
\begin{equation}
A\leftrightarrow\hat A\col\quad
Y\leftrightarrow Y^\dagger\col\quad
\psi\leftrightarrow\psi^\dagger\col\quad
c\leftrightarrow\hat c\col\quad
c^\star\leftrightarrow\hat c^\star\col
\end{equation}
and it exchanges 
the colour loops of both gauge groups, i.e.\ it exchanges
$\lambda\leftrightarrow \hat\lambda$. We also have to consider 
several sign changes due to the exchange of certain vertices and of the 
propagators of the two gauge fields. By 
$P_x$ and $V_x$ we denote the multiplicities with which the corresponding 
propagator or vertex of type $x$ appears in the graph.
A shift by one side then transforms a Feynman graph as
\begin{equation}
c(\lambda,\hat\lambda)\pthree{a_1}{\dots}{a_m}\to(-1)^{P_{A^2}+V_{A^3}+V_{A\psi^2}+V_{Y^2\psi^2}}c(\hat\lambda,\lambda)\pthree{a_1+1}{\dots}{a_m+1}
\pnt
\end{equation} 
Furthermore, the reflection of a given non reflection-symmetric Feynman graph 
contributes to the perturbation series. A reflection exchanges $A\leftrightarrow\hat A$ and it changes also the sign of all loop momenta.
A reflection hence transforms a Feynman graph as
\begin{equation}
c(\lambda,\hat\lambda)\pthree{a_1}{\dots}{a_m}\to(-1)^{P_{A^2}+V_{AY^2}+V_{A\psi^2}+V_{Y^2\psi^2}}c(\hat\lambda,\lambda)\pthree{-a_1}{\dots}{-a_m}
\pnt
\end{equation}

\end{fmffile}


\footnotesize
\bibliographystyle{JHEP}
\bibliography{references}

\end{document}
